\documentclass[pdflatex,sn-basic]{sn-jnl}
\usepackage{graphicx}%
\usepackage{multirow}%
\usepackage{amsmath,amssymb,amsfonts}%
\usepackage{amsthm}%
\usepackage{mathrsfs}%
\usepackage[title]{appendix}%
\usepackage{xcolor}%
\usepackage{textcomp}%
\usepackage{manyfoot}%
\usepackage{booktabs}%
\usepackage{algorithm}%
\usepackage{algorithmicx}%
\usepackage{algpseudocode}%
\usepackage{listings}%
\usepackage{comment}
\usepackage{makecell}
\usepackage[T1]{fontenc}

\def\aap{Astron. Astrophys.}                
\def\aj{Astron. J.}

\def\apj{Astrophys. J.}
\def\apjl{Astrophys. J. Lett.}
\def\apjs{Astrophys. J. Suppl. Ser.}

\def\mnras{Mon. Not. R. Astron. Soc.}

\def\nat{Nature}

\def\pasp{Publ. Astron. Soc. Pac.}

\def\araa{Annu. Rev. Astron. Astrophys.}

\def\prd{Phys. Rev. D}

\newcommand{\Msun}{\ensuremath{M_{\odot}}}

\theoremstyle{thmstyleone}%
\theoremstyle{thmstyletwo}%

\theoremstyle{thmstylethree}%

\begin{document}

\title[Disc Instabilities]{Non-Stationary Discs and Instabilities}


\author*[1]{\fnm{Omer} \sur{Blaes}}\email{blaes@physics.ucsb.edu}
\equalcont{These authors contributed equally to this work.}

\author[2]{\fnm{Yan-Fei} \sur{Jiang}}\email{yjiang@flatironinstitute.org}
\equalcont{These authors contributed equally to this work.}

\author[3,4]{\fnm{Jean-Pierre} \sur{Lasota}}\email{lasota@iap.fr}
\equalcont{These authors contributed equally to this work.}

\author*[5]{\fnm{Galina} \sur{Lipunova}}\email{gvlipunova@gmail.com}
\equalcont{These authors contributed equally to this work.}

\affil*[1]{\orgdiv{Physics Department}, \orgname{University of California at Santa Barbara}, \orgaddress{\city{Santa Barbara}, \state{CA}, \postcode{93106}, \country{USA}}}

\affil[2]{\orgdiv{Center for Computational Astrophysics}, \orgname{Flatiron Institute}, \orgaddress{\city{New York}, \state{NY}, \postcode{10010}, \country{USA}}}

\affil[3]{\orgdiv{Nicolaus Copernicus Astronomical Center}, \orgname{Polish Academy of Sciences}, \orgaddress{\street{Bartycka 18}, \city{Warsaw}, \postcode{00-716}, \country{Poland}}}

\affil[4]{\orgdiv{Institut d'Astrophysique de Paris}, \orgname{CNRS et Sorbonne Universit\'e,
           UMR 7095}, \orgaddress{\street{98bis Bd. Arago}, \city{Paris}, \postcode{75014},  \country{France}}}

\affil[5]{
\orgname{Max Planck Institute for Radio Astronomy}, \orgaddress{\street{Auf dem Hügel 69}, \city{Bonn}, \postcode{53121}, 
\country{Germany}}}



\abstract{We review our current knowledge of thermal and viscous instabilities in accretion discs around compact objects.  We begin with classical disc models based on analytic viscosity prescriptions, discussing physical uncertainties and exploring time-dependent solutions of disc evolution.  We also review the ionization instability responsible for outbursting dwarf nova and X-ray binary systems, including some detailed comparisons between alpha-based models and the observed characteristics of these systems.  We then review modern theoretical work based on ideas around angular momentum transport mediated by magnetic fields, focusing in particular on knowledge gained through local and global computer simulations of MHD processes in discs.  We discuss how magnetohydrodynamics (MHD) may alter our understanding of outbursts in white dwarf and X-ray binary systems.  Finally, we turn to the putative thermal/viscous instabilities that were predicted to exist in the inner, radiation pressure-dominated regions of black hole and neutron star discs, in apparent contradiction to the observed stability of the high/soft state in black hole X-ray binaries.}




\maketitle

\section{Introduction}\label{sec:intro}




The seminal 1973 paper by \citeauthor{sha73} laid out a basic model for the radial structure of stationary, geometrically thin accretion discs around black holes.  It did so by assuming that the vertically-averaged angular momentum-transporting stress $w$ in the disc was approximately proportional to the vertically-averaged total thermal pressure $P$ in the disc\footnote{The exact definition used by \citeauthor{sha73} (equation 1.2 of their paper) actually defined the stress as $\alpha$ times the density times the square of the sound speed, the latter being defined in terms of the total thermal energy density and the density.  Apart from factors of order unity, this is equivalent to Eq.~(\ref{eq:alpha_prescription}).}:
\begin{equation}
   w=\alpha P\, .
   \label{eq:alpha_prescription}
\end{equation}
This ansatz (the alpha prescription) became the basis for virtually all analytic models of accretion discs since then, including those of accretion discs around objects other than black holes.

One year after the original Shakura \& Sunyaev paper, it was pointed out by \citet{lig74} that the radiation pressure and Thomson scattering dominated inner zone was unstable.  Assuming vertical hydrostatic and thermal equilibrium, the standard alpha prescription leads to an inverse relationship between vertically integrated stress $W$ and surface mass density $\Sigma$ in the disc, which in turn results in an effectively negative radial mass diffusion coefficient.  Fluctuations in surface mass density would then be amplified, likely causing the disc to break up into rings.

In another seminal paper, \citet{sha76} followed this up with an analysis that also accounted for time variations in the thermal balance of the disc, and found that the instability discovered by Lightman \& Eardley was part of a more complicated set of instabilities that exist in the inner disc region.  At long radial wavelengths $\Lambda$, these instabilities separate into two branches:  the anti-diffusion ``viscous'' instability of Lightman \& Eardley, with growth rate $\sim\alpha(H/\Lambda)^2\Omega$, and a faster thermal instability with growth rate $\sim\alpha\Omega$.  Here $H$ is the local disc vertical thickness, and $\Omega$ is the local orbital angular velocity.  The thermal instability arises in the standard alpha prescription because the local heating rate per unit area $Q^+\propto\alpha T^8/\Sigma$, while the cooling rate $Q^-\propto T^4/(\kappa\Sigma)$.  The temperature dependence of the heating rate at fixed surface mass density is therefore much steeper than that of the cooling rate if Thomson opacity dominates the opacity $\kappa$, causing runaway heating or cooling.

Unfortunately, it has never been clear whether these instabilities have a physical reality, or are merely an artifact of the assumptions behind the alpha-prescription.  Simple modifications to the alpha prescription can in fact stabilize the radiation pressure dominated inner regions of black hole accretion discs \citep{pir78}.  For example, rapid photon diffusion might decouple turbulent fluctuations from the radiation pressure, so that the stress might scale in proportion to just the pressure in the gas \citep[e.g.][]{lig74,sak81,ste84}. Or perhaps the speeds of turbulent eddies would be limited to the gas sound speed but still be of a size that reflects the radiation pressure scale height, leading to a stress that is proportional to the geometric mean of gas and radiation pressure~\citep[e.g.][]{taa84}.  Both of these prescriptions would stabilize the inner disc.  Winds can also help stabilize the disc~\citep{pir77}.  Moreover, in the case of discs around supermassive black holes, the opacity might not be dominated by electron scattering, and this could also stabilize the inner regions of the disc~\citep{pri76}.  Finally, even if the stress does scale with the total pressure in the disc, that pressure might be dominated by temperature-independent magnetic pressure rather than thermal pressure~\citep{shi90}, with dissipation happening above the disc~\citep{fie93}, within the disc \citep{par03,beg07,oda09}, or a combination of the two \citep{beg15}.

Still another possibility that avoids the thermal and viscous instabilities of the inner regions of black hole accretion discs is that the accretion power that is dissipated at each radius is not radiated away locally, but is instead advected inward.  This generates a whole new family of advection-dominated accretion disc solutions based on the Shakura-Sunyaev alpha-prescription that are thermally and viscously stable \citep{abr88,nar94,abr95,nar95a,nar95b}.  The local topology of these solutions and its relation to non-advective models is very nicely summarized in the short paper by \citet{che95}.  Such models have been applied to quiescent, low/hard, and intermediate states \citep{esi97,esi98} and to near-Eddington accretion states \citep{str11} \citep[so-called ``slim'' accretion discs,][]{abr88} of black hole X-ray binaries.  However, the high/soft state is still generally modeled as a geometrically thin disc with little advection \citep{esi97,str11,dav06}, and the extremely low variability of this state suggests that it is stable \citep{gie04}.  Advective models have also been applied to low luminosity \citep{nar98} and high luminosity \citep{szu96} accretion onto supermassive black holes.

In this paper we review our current understanding of the nature of thermally and viscously unstable behavior in accretion discs around compact objects.  In \S\ref{sec2}, we first review analytic solutions of time-dependent evolution of discs with analytic viscous stress prescriptions.   Then in \S\ref{sec3} we briefly review the various ways in which angular momentum transport in sufficiently ionized discs is likely mediated by magnetic fields.  In marked contrast to the original discovery of thermal and viscous instabilities in black hole accretion discs, these instabilities actually appear to play a fundamental role in driving observed outburst behavior in cataclysmic variables and the outer discs of low-mass X-ray binaries.  We review the basics of the disc instability model as applied to these systems in \S\ref{sec4}.  We then return to the putative instabilities in the inner regions of black hole accretion discs in \S\ref{sec5}.  We conclude with a list of important, outstanding questions and suggestions for future research in \S\ref{sec6}.

We do not discuss models of collimated jets in this review, nor do we discuss so-called MAD (magnetically arrested disc) flows which are an important aspect of the jet problem.  These topics are thoroughly discussed elsewhere.  
We will discuss magnetocentrifugal winds, however, as they may play an important role in the operation of thermal and viscous instabilities in discs.

\section{Analytic Solutions for Time-Dependent Discs}\label{sec2}


%

\long\def\COMMENT#1{\relax}
\newcommand{\ar}{a_\mathrm{r}}
\newcommand{\boldnabla}{\mbox{\boldmath$\nabla$}}
\newcommand{\Bf}{B_\mathrm{ff}}
\newcommand{\br}{b_\mathrm{r}}
\newcommand{\den}{\mathrm{d}}
\newcommand{\de}{\partial}
\newcommand{\EBV}{E(\mbox{B}-\mbox{V})}
\newcommand{\Mus}{\mbox{GS\,1124--683\protect}}
\newcommand{\ho}{h_0}
\newcommand{\hk}{h_{\scriptscriptstyle{\mathrm{K}}}}
\newcommand{\kapf}{\varkappa_{\scriptscriptstyle{\mathrm{ff}}}}
\newcommand{\kapt}{\varkappa_{\scriptscriptstyle{\mathrm{T}}}}
\newcommand{\kappaT}{\kappa_{\scriptscriptstyle{\mathrm{T}}}}
\newcommand{\Led}{L_{\mathrm{Edd}}}
\newcommand{\ls}{\left(}
\def\Mon{{\hbox{A\,0620--00}}}{}%
\newcommand{\mx}{m_\mathrm{x}}
\newcommand{\NH}{N(\mathrm{HI})}
\newcommand{\nutc}{\nu_\mathrm{t}^\mathrm{c}}
\newcommand{\nut}{\nu}
\newcommand{\omegak}{\Omega_{\scriptscriptstyle{\mathrm{K}}}}
\newcommand{\pc}{P_\mathrm{c}}
\newcommand{\ps}{\right)}
\newcommand{\Qo}{Q_0}
\newcommand{\omegad}{\Omega}
\newcommand{\rg}{r_{\mathrm{g}}}
\newcommand{\Rg}{R_{\mathrm{g}}}
\newcommand{\Rs}{R_{\mathrm{g}}}
\newcommand{\rhoc}{\rho_\mathrm{c}}
\newcommand{\rin}{r_{\mathrm{in}}}
\newcommand{\rout}{r_\mathrm{out}}
\newcommand{\risco}{r_{{\scriptscriptstyle{\mathrm{ISCO}}}}}
\newcommand{\sigmasb}{\sigma_{{\scriptscriptstyle{\mathrm{SB}}}}}
\newcommand{\sigmat}{\sigma_{\scriptscriptstyle{\mathrm{T}}}}
\newcommand{\sigmo}{\Sigma}
\newcommand{\slfrac}[2]{\left.#1\middle/#2\right.}
\newcommand{\sun}{\odot}
\newcommand{\shsc}{\cite{shakura-sunyaev1973}}
\newcommand{\tauf}{\tau_\mathrm{ff}}
\newcommand{\taut}{\tau_{\scriptscriptstyle{\mathrm{T}}}}
\newcommand{\tc}{T_\mathrm{c}}
\newcommand{\tobs}{\Delta t_\mathrm{ob}}
\newcommand{\teff}{T_\mathrm{eff}}
\newcommand{\Zo}{Z_0}

\newcommand{\added}[1]{\textcolor{blue}{#1}}

\def\aap{A\&A{ }}%
\def\apjl{ApJ{ }}%
\def\apj{ApJ{ }}%
\def\aj{AJ{ }}%
\def\apjs{ApJS{ }}%
\def\mnras{MNRAS{ }}%
\def\apss{Ap\&SS{ }}%
\def\araa{ARA\&A{ }}%
\def\prd{Phys.~Rev.~D{ }}%
\def\nat{Nature{ }}%
\def\pasp{PASP{ }}%
\def\azh{Astronomy Reports{ }}%
\def\pasj{PASJ{ }}%
\def\na{NewA{}}%
\def\actaa{Acta Astr.{}}%


\newcommand*{\oline}[1]{\overline{#1\mathstrut}}


In this section we discuss a variety of analytic solutions for the time-dependent behavior of accretion discs.  Construction of such solutions requires various simplifying assumptions, particularly in the stress prescription (or equivalently, an effective viscosity).  Although real accretion disks may not fully satisfy these assumptions, such solutions are nevertheless valuable in understanding the behavior of more complex simulations with more realistic physics, particularly MHD.  Moreover, the present-day limitations of computational power mean that analytic solutions are still the only way to model the evolution of discs on viscous time scales.  Eventually, one hopes that simulation data can be used to improve upon the assumptions that are used in constructing these solutions.

The vertically integrated equations of mass and angular momentum conservation can be combined to yield
an equation governing the evolution of the surface density $\sigmo$ (the density, integrated over the disc thickness) for a disc with an arbitrary 
radial profile of angular velocity $\omegad(r)$:
\begin{equation}
\frac{\partial \sigmo} {\partial t} =  
\frac{1}{r}\, \frac{\partial}{\partial r} \left[ 
\left(\frac{\partial \, \omegad  r^2}{\partial r} \right)^{-1} \,
\frac {\partial}{\partial r} \left(-r^3 \,\frac{\partial \omegad}{\partial r}\, \nu \, \sigmo\right)
\right]\,  
\label{eq.time-dep-acc1}
\end{equation}
\citep[e.g.][]{lig74,lyn-pri1974,sha76,Pringle1981,lin-pringle1987,shakura_etal2018}, where $\nu$ is the coefficient of kinematic viscosity.
The local effective kinematic viscosity in the discs can be associated with various mechanisms that ensure the angular momentum transfer:  hydrodynamic~\citep[see][for a review]{Lesur+2023}, magnetorotational turbulence~\citep{bal91}, convection-driven turbulence~\citep{Lin-Papaloizou1980} and self-gravity instabilities~\citep{Paczynski1978}.
In weakly-magnetized accretion discs, where shear and substantial ionization are present, magnetorotational (MRI) turbulence is thought to play a major role, see \S\ref{sec:MRI}.

This equation serves as a key element in the analysis of the disc's behavior over time, capturing everything from the instabilities of accretion discs --- where,  for example, when used alongside the energy conservation equation, it reveals the nature of viscous-thermal instability --- to the long-term evolution.

Mathematically, Eq.~\eqref{eq.time-dep-acc1} is a parabolic equation of the second order in partial derivatives, that is, a diffusion equation for $\Sigma$, which generally  is 
nonlinear  because the viscosity can depend on the surface density~\citep{lig74}. In the case of Keplerian orbits, the equation can be written much more simply.  
In particular, if it is written in terms of the $z$-integrated  viscous torque $F = -r \, 2\pi r \Sigma \nu r \,{\partial \omegad}/{\partial r}$,  which is a couple exerted on a ring of material by the disc interior to $r$ and becomes $ F=   3\,\pi\, h\, \nu\,\sigmo$, and the specific Keplerian angular momentum $h = \omegak\, r^2 = \sqrt{G M r}$.
The disc-evolution equation can then be written as~\citep{filipov1984,lyub-shak1987}:
 \begin{equation}
\frac{\partial F}{\partial t}=
D\,\frac{F^m}{h^n}\,\frac{\partial^2F}{\partial h^2}\, .
\label{eq.nonlin-diff}
\end{equation}
In this nonlinear diffusion equation $m$ and $n$ are the scaling exponents and $D$ is the diffusion coefficient, which follows from imposing local thermal equilibrium:
\begin{equation}
 D = \frac{(GM)^2\, F^{1-m}}{4\,\pi\,(1-m)\,\sigmo\, h^{3-n} }\, . 
\label{eq.SigDF}
 \end{equation}
Eq.~\eqref{eq.nonlin-diff} follows the view used by \citet{lyn-pri1974} with their $g(h,t)=F(h,t)$. 
It assumes a positive value of the viscous torque $F$ for a standard disc, while the net viscous torque  (that is, the difference of the outer and inner torques, applied to a ring) is negative in the disc.  This convention is seen in a number of works and 
is 
consistent with the frequent situation when the parameter $m$ is non-integer.

To relate $D$ to the kinematic viscosity $\nu$, it is sufficient to express the latter as a product of power law functions 
$$
\nu = \nu_0\,\Sigma^a\, r^b\, ,$$  
and, accordingly, the viscous torque as 
\begin{equation} F = 3\,\pi\, h\, \nu\,\Sigma = 3\,\pi\, h\, \nu_0\,\Sigma^{a+1}\, r^b\, .
\label{eq.Fab}
\end{equation}
Here $\nu_0$ is a dimensional normalization factor. Substitution of \eqref{eq.Fab} into the l.h.s. of
$$
\frac{\partial \sigmo} {\partial t} =  \frac{(GM)^2}{4\pi h^3}\, \frac{\partial^2F}{\partial h^2}\, 
$$
yields
\begin{equation}
 D = \frac{a+1}{2}\, (G\,M)^2\, \left(\frac 32\,
\frac{\nu_0}{(2\,\pi)^a\,(G\,M)^b}\right)^{1/(a+1)}~,
\label{eq.Dnice}
\end{equation}
and the dimensionless parameters  $m$ and $n$:
\begin{equation}
    m=\frac{a}{a+1}\, , \qquad n = \frac{3a+2-2b}{a+1}\, .
    \label{eq.mnab}
\end{equation}
The normalization $\nu_0$  can be predicted using \eqref{eq.Dnice} if the vertical structure of the disc is calculated, so that $D$ is found via Eq.~\eqref{eq.SigDF}.

In the variables $F$ and $h$, the equation of angular momentum transfer takes the form:
\begin{equation}
\dot M (r,t)= \frac{\partial F}{\partial h}\, .
\label{eq.acc-rate2}
\end{equation}
Therefore, the disc evolution equation in form \eqref{eq.nonlin-diff} has the advantage of allowing one to impose transparent explicit boundary conditions. Also, the flux radiated from one surface of the disc can be written in a unified form applicable to a general case of accretion/decretion discs~\citep{rafikov2013}: 
\begin{equation}
Q_{\rm vis} = \frac{3}{8\pi} \frac{F\omegak}{r^2}\, .
\label{eq.Qvis_stationary}
\end{equation}

The characteristics of viscosity and the vertical structure of the disc determine the parameters $a$ and $b$, as well as $D$, $m$ and $n$; see Table~\ref{tab.indexes} below~\citep[see, e.g.,][]{shakura_etal2018}.  Usually the viscosity depends on the disc physical conditions and thus on the accretion rate in the disc. For example, the anti-diffusion viscous instability found by \citet{lig74} in the inner radiation-pressure dominated and Thomson scattering zone corresponds to the negative effective diffusion coefficient $D$ with  $a=-2$ in Eq.~\eqref{eq.Fab}\footnote{The solution to the angular momentum balance equation \eqref{eq.acc-rate2}  is $\dot M \, h \,f(r)= F$, where $f(r)$ is some function. In the $\alpha$-disc, the viscous torque  $F = 2\pi r^2 \, 2z\, \alpha\,P$.  Substituting the  pressure from the hydrostatic balance $P \sim \Sigma \,z\omegak^2$ in the latter, one obtains $\dot M\, f(r) \sim  4\pi\,\alpha z^2 \, \Sigma\omegak$.
In the radiation pressure dominated zone with Thomson opacity, the half-thickness of the disc $z\sim 3\dot M \sigma_{\rm T}/(8\pi c) $, which yields
 $\Sigma \propto \dot M^{-1}$.  Equating the viscous torque to the expression in \eqref{eq.Fab}, we find the dependence for the kinematic turbulent viscosity:
 $\nu \propto \Sigma^{-2}$.}.

At large distances around normal or compact stars, where gas pressure dominates and/or opacity is determined by absorption processes in the ionized matter, $D$ is positive and a stable long-term evolution is possible. 
For $\alpha$-discs, the diffusion coefficient $D$ depends only weakly on the opacity: as a power law  with an index of ${1/5}$ or ${1/10}$ for Thomson or Kramers opacities, respectively. This mitigates the effect of the uncertainty associated with the dependence of the real opacity on the disc parameters. Consequently, $D$ can be regarded as constant in the equation of non-stationary accretion~(\ref{eq.nonlin-diff}).

A method to determine the long-term evolution from the equation of non-stationary accretion \eqref{eq.nonlin-diff}  depends on the form of the turbulent viscosity coefficient $\nut=\nut(r,\sigmo)$.  If the viscosity $\nut$ is a function of the radius only and does not depend on the surface density (i.e. does not depend on time), then $F$ depends linearly on the surface density $\sigmo$.  For such cases,  when $m=0$ and $a=0$, 
Eq.~\eqref{eq.nonlin-diff} becomes a linear diffusion equation.
Note that in this case the characteristic viscous time scale $\tau_\mathrm{vis}\sim r^2/\nut$ is constant in time.
In a more general case, $\nut$ also depends on the surface density. In particular, if we consider discs with $\alpha$-viscosity, we can represent $\nut$ as a power-law function of $\Sigma $ and $r$.

A general solution to the linear diffusion equation can be determined through an eigenfunction expansion. 
The method of superposition enables the formulation of a solution that complies with the specified initial or boundary conditions.
In the case of a nonlinear diffusion equation, separation of variables provides solutions only for a certain class of problems. 
An understanding of the physical properties of problems can, in certain situations, enable the derivation of self-similar solutions. 
A self-similar solution of the nonlinear diffusion equation describes the evolution when enough time has passed so that the initial state is forgotten.

Here we review some known analytic solutions of Eq.~\eqref{eq.time-dep-acc1} with different boundary conditions. Regardless of the details of an initial supply of matter, if the expansion of a disc is unrestricted, it will eventually enter a phase 
when its outer parts acquire high angular momentum and move farther and farther away from the center. 
Then the discs, even if they are accretionary near the center, are decretionary at large distances (\S\ref{s.linear_freeexp} and \S\ref{s.non-lin-equation-freely}).
In binary systems, the gravitational interaction with the companion star prevents such unrestricted expansion of a disc. This leads to different solutions compared to freely expanding discs (\S\ref{s.lin-equation-bounded} and \S\ref{s.non-lin-equation-bounded}).

In the presence of external torques, Eq.~\eqref{eq.time-dep-acc1} is not valid, strictly speaking.  If the disc rotates around a binary or in a binary, the tidal interaction exchanges the angular momentum between the disc and the binary orbital motion. In both cases, it is common to assume that the tidal torque is accumulated in a narrow disc ring, the inner or the outer one~\citep{pringle1991}. Therefore, the influence of a binary on a disc can be represented as an effective boundary condition for Eq.~\eqref{eq.time-dep-acc1}.

When accreting onto a black hole, the disc is often assumed to have zero torque at the inner boundary~\citep[standard disc by][]{sha73}. The other situation --- where there is no accretion through the inner boundary, but there is a finite torque at the inner boundary $\rin$ --- is thought to occur for rapidly rotating magnetized neutron stars or discs around binary systems. 
 \citet{rafikov2016} showed that there is indeed a whole range of possibilities for the internal boundary condition, with generally non-zero central viscous stress and accretion rates varying from positive to negative values (e.g. if the disc is gaining mass from a central Be-star); see also \citet{Nixon-Pringle2021}.

The tidal forces of a companion star confine the disc within the Roche lobe of the accretor~\citep{pap-pri1977,paczynski1977,ich-osa1994,hameury-lasota2005}.
Significant perturbations near the outer disc radius lead to the formation and dissipation of shocks and the loss of angular momentum from the disc, which is transferred to the orbital motion. 
\citet{pap-pri1977} showed that the tidal {truncation} radius is on average $\sim 0.9$ times that of the Roche lobe. 
This radius is close to that of the last non-intersecting periodic orbit calculated for a three-body problem~\citep{paczynski1977}.
Numerical calculations have shown that the tidal stress tensor is only important in a rather narrow ring close to the outer radius. 
Therefore, one can choose to not study this region in detail by considering tidal interactions to occur in a zero-width region and assuming that $r_{\rm out}$ is constant.

\subsection{Linear equation, freely expanding disc}\label{s.linear_freeexp}

In 1952, \citeauthor{lust1952} found particular solutions to the  linear equation of viscous accretion, proposed by his teacher von Weizs\"acker~(\citeyear{weizsaecker1948}), and described the principles of constructing a general solution to the evolution of discs with either infinite or finite geometrical extent.

\begin{figure}
    \centering
\includegraphics[width=0.49\linewidth]{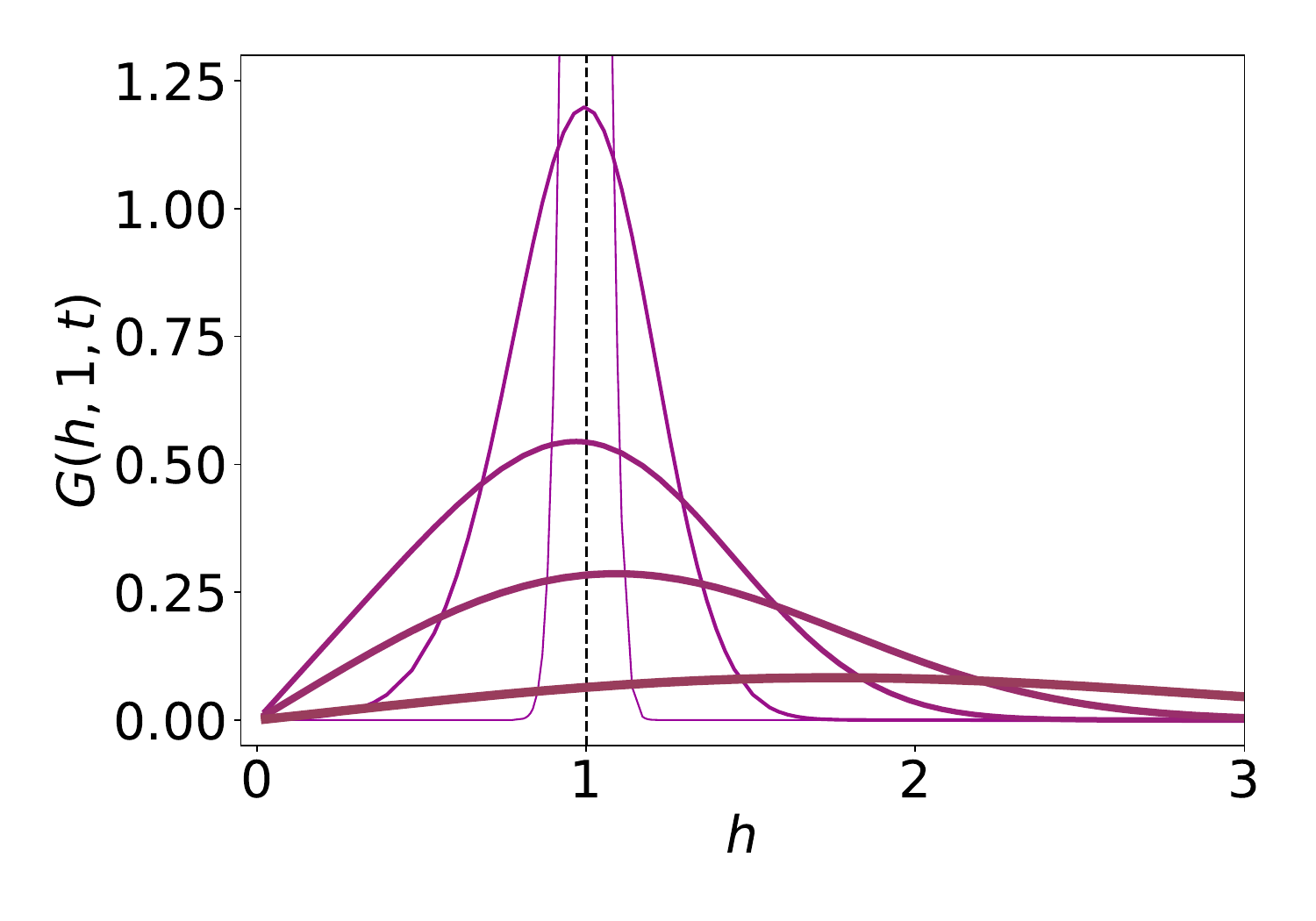}
\includegraphics[width=0.49\linewidth]{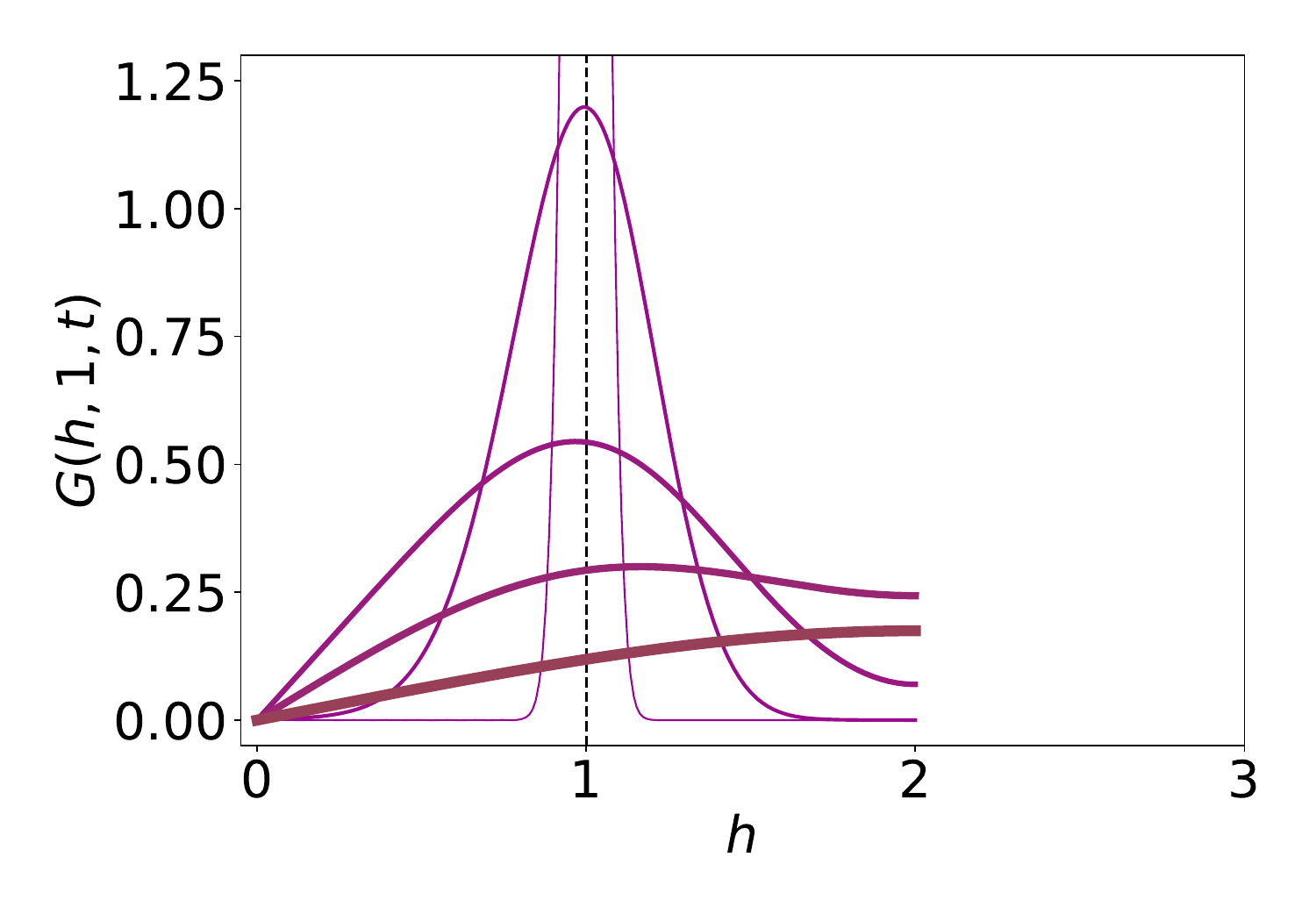}
    \caption{Green function $G(h,1,t)$ versus specific angular momentum $h$ for freely expanding disc (left) and for disc with the outer radius fixed so that $h_{\rm out}=2$.  Thicker curves correspond to later times.  
    For the bounded disc, the condition of zero accretion rate is applied at the outer radius.
}
\label{fig:green_functions}
\end{figure}

\citet{lyn-pri1974} used a method of superposition of particular solutions to the equation of viscous evolution and found general solutions for two types of boundary conditions at the inner boundary: $F=0$ (zero viscous torque at the inner radius, applicable, perhaps, to the case of accretion onto a black hole)  and $\partial F/\partial h = 0$ (zero accretion rate). 
Given the initial distribution $F_0(h)$,  the solution to the linear differential equation is a linear integral transform: 
$$
F(h,t) = \int\limits_0^\infty  G(h,h_1, t) \, F_0(h_1)\, \mathrm{d}
h_1\,,
$$
where $G$ is the Green's function of  Eq.~\eqref{eq.nonlin-diff} with $m=0$ ($a=0$). 
The problem of solving a linear differential  equation with boundary conditions  can be viewed as a linear system with an input signal $F_0(h_1)$ and an output signal $F(h,t)$.
The Green's function then acts as the system's response to a delta impulse input signal $F_0(h_1)=\delta(h_1-h_{\rm s})$. The Green's function itself is obtained as the solution to a boundary value problem with the continuum spectrum of eigenvalues and the Dirac delta function as the initial condition. As a result, $G$ is an integral containing Bessel functions of non-integer order that can be taken analytically by a Hankel transform~\citep[see \S 2.4 in][]{lyn-pri1974}. An example of the Green's function $G(h,1,t)$ is shown in Fig.~\ref{fig:green_functions}.

With the help of the Green's functions it is possible to find $F$, $\Sigma$, $\dot M$ at any moment in time and at any point for arbitrary initial conditions. For example, the accretion rate can be expressed from its corresponding Green's function:
$$
\dot M(h,t) = \int\limits_0^\infty  G_{\dot M}(h,h_1, t) \, F_0(h_1)\, \mathrm{d}
h_1/h_{\rm s}\, .
$$
In the case of the accretion onto a black hole from a viscously spreading ring, the accretion rate at the inner disc boundary has the explicit analytic form \citep{lyn-pri1974}: 
\begin{equation}
\dot M_\mathrm{in} (t) = \dot M_{\mathrm{in}, \mathrm{max}}  \left(
\frac{\tau_\mathrm{pl}}{t}\right)^{1+l}\, e^{-\tau_e/t}\, ,
\label{eq.LBP_solution_Mdot}    
\end{equation}
where the characteristic time scale for the exponential growth $\tau_e$ and power-law decline \index{viscous disc evolution!power-law decay} 
$\tau_\mathrm{pl}$ depend on the initial size and viscosity $\nu = \nu_0 \, r^b$ in the disc:
\begin{equation}
    \tau_e = \frac{\kappa^2\, h_\mathrm{s}^{1/l}}{4} =  \frac{1+l}{e}\,
\tau_\mathrm{pl}\,  , \quad \mathrm{where} \quad \frac{1}{2\,l} = 2-b\, \quad \mathrm{and} \quad \kappa^2 = \frac{16 \, l^2}
{3\nu_0\,(G\,M)^{1/2l}}\, .
\label{eq.lin_evol_pars}
\end{equation}
It is assumed that the  ring with mass $M_\mathrm{disc}$ was originally  located at $r_s = h_s^2/GM$. It is  easy to see  from \eqref{eq.nonlin-diff} and \eqref{eq.Dnice} that $D = 4 \,({l}/{\kappa})^2$ for $m=0$. Here, $l>0$ and $b<2$.

The accretion rate reaches its peak value
\begin{equation}
\dot M_{\mathrm{in}, \mathrm{max}} =  \frac{M_\mathrm{disc}}{t_\mathrm{max}} \, 
\frac{(1+l)^l}{e^{1+l}\, \Gamma(l)}\, 
\label{eq:LBP_mdot}
\end{equation}
at the time~\citep{lipunova2015} 
\begin{equation}
t_\mathrm{max} = \frac{4\,l^2}{3\,(1+l)} \,  \frac{r_\mathrm{s}^2}{\nu_\mathrm{s}} = 
\frac{\tau_\mathrm{pl} }{e}\, . 
\label{eq:ap.dotmmax_snfinite}
\end{equation}
Here $\nu_\mathrm{s} = \nu_0 \,r_\mathrm{s}^b$. Assuming the values of $l$, original radius $r_s$, and time $t_\mathrm{max}$ are known, the normalization factor $\nu_0$ can be obtained.

At later times after the peak, the accretion rate through the inner boundary declines as a power law $\dot M \propto t^{-1-l} $, where the parameter $l<1$. 
Each solution for $\Sigma$-independent viscosity extends to infinity for any $t$. 

\citet{pringle1991} found the Green's functions for an `external' disc with zero accretion rate at $\rin>0$ and for $\nu \propto r$. In this particular case of $b=1$, the Bessel functions reduce to functions of the form 
$\sin x/\sqrt x$ and $\cos x/\sqrt x$.
\citet{tanaka2011} found Green’s function solutions for arbitrary $b$ that satisfy either a zero-torque or zero-flux condition when the inner boundary of the disc is located at a finite radius $\rin>0$.

For zero accretion rate and a finite torque at the inner boundary, $\Sigma$ and $F$ vary with time as $\propto t^{-1+l}$. Keeping in mind that the bolometric flux varies according to \eqref{eq.Qvis_stationary}, we see that the luminosity of such discs decreases more slowly with time than that of discs with vanishing inner stress~\citep[see also][where the numerical solution for a disc around a Kerr black hole and non-vanishing stress is considered in the context of TDE observations]{Balbus-Mummery2018}. Near the center, the torque $F$ develops a flat profile, the amplitude of which rises at first and then decreases with time. \citet{rafikov2016}
found general late-time asymptotics for non-zero accretion rate and non-zero inner torque (for a general case with $a\neq0$ too, see \S\ref{s.non-lin-equation-freely}). It was shown that suppression of the central accretion rate always requires stronger central torque. Conversely, the central torque on the disc suppresses mass accretion on the object or even reverses it to an outflow, as in decretion discs around Be-stars.

A general property of a linear differential equation is that the superposition of its solutions yields another valid solution.
Using the $\nu\propto r$ case as an example, \citet{Nixon-Pringle2021} showed how to obtain the Green's function in the case of a general  finite torque boundary condition at the inner boundary
\begin{equation}
     \Big[F - f \,\frac{\partial F}{\partial h}\Big] _{\rin} = 0\, , 
     \label{eq.general_inner_cond}
\end{equation}
where $f\geq 0$ is an arbitrary constant.

 \citet{Balbus2017} has derived a form of the thin disc diffusion equation that is valid for both the Schwarzschild and Kerr geometries. The Green's function solutions can be calculated  in terms of quadratures by using a combination of WKB technique, local analysis and matched asymptotic expansions. \citet{Mummery2023} computed analytically the leading order Green’s function solutions (for $F$, $\Sigma$ and $\dot M$)\footnote{In \citet{Mummery2023},  $\mu = b-1/2$ and $\alpha=1-b/2$.} of the general relativistic thin disc equations for a vanishing stress at the innermost stable
circular orbit. It turns out that the peak mass accretion rate across the innermost stable circular orbit (ISCO) is lower for more rapidly rotating Kerr black holes. When the accretion efficiency dependence on the spin is taken into account, the rest-frame luminosity of the more rapidly spinning
black holes is the largest.

\subsection{Linear equation, bounded disc}
\label{s.lin-equation-bounded}

Similarly to the case of a freely expanding disc, the method of superposition of particular solutions to Eq.~\eqref{eq.nonlin-diff} can be applied.
The difference is that, for a bounded disc with $\rout = $const, the corresponding eigenvalue problem has a purely discrete spectrum.
Therefore, the Green's function is not an integral, but a linear combination of particular solutions satisfying the specific boundary conditions.
A particular Green's function for a finite disc was constructed by \citet{wood_etal2001}. 
The full Green's function, which can be used for an arbitrary initial density distribution for the two types of inner boundary conditions, has been constructed by \citet{lipunova2015} for $\rin=0$ (Fig.~\ref{fig:green_functions}, the right panel); see also \citet{Mushtukov+2019} for $\rin>0$. 

\begin{table*}
 \centering
 \scriptsize
  \caption[Parameters of the Green function for a non-stationary disc]
  {Parameters of the time-dependent solution with index $a=0$ in viscosity $\nu \propto \Sigma^a \,r^b $. The ring starts at $r_{\rm s}$ and evolves into the disc with zero inner torque.  If the disc  expands freely then the central accretion rate decays as $ \propto t^{-1-l} $, see Eq.~\eqref{eq.LBP_solution_Mdot}. If the disc has a fixed outer boundary, $\dot M$ decays exponentially, see Eq.~\eqref{eq:texp}. For such a disc the peak time is similar to the case of a freely expanding case ~\eqref{eq:ap.dotmmax_snfinite}, for which it is the exact value.
  Parameter $k_1$ is  an eigenvalue  for the solution in the interval ($0,r_{\rm out}$).
  Solutions with $a=0$ and specific $b$ approximate the $\alpha$-disc evolution on time scales  of the order of a couple of viscous timescales.  Different opacity laws (the Rosseland mean) are indicated for $\alpha$-discs.
\label{tab:zeroes}}
  \begin{tabular}{@{}lccccc@{}}
\hline
{Disc type} & $b$ & $l$ & {$k_1$} & {$t_\mathrm{max} (r_\mathrm{s}^2/\nu_\mathrm{s})^{-1}$} & 
{$t_\mathrm{exp} (r_\mathrm{out}^2/\nu_\mathrm{out})^{-1}$} \\
\hline
$\nu=const$ & 0     & $1/4$ & 1.06 & $1/15$ & $ 0.30$ \\ [0.3em]
\makecell{Disc with $z/r=$const,\\e.g., ADAF~~~~~~~~~~} & $1/2$ & $1/3$ & 1.24 & $1/9$ &$ 0.38 $ \\ [0.7em]
$\alpha-$disc, $\kappa=$const & $3/5$ & $5/14$ & 1.29  & $ 0.125$  & $ 0.41 $ \\  [0.3em]
$\alpha-$disc, $\kappa \propto \rho T^{-7/2}$ & $3/4$ & $2/5$ &  1.38     & $ 0.152$  &$ 0.45$ \\   [0.3em]
$\alpha-$disc, $\kappa \propto \rho T^{-5/2}$ &  $2/3$ & $3/8$ &   1.33    &  $3/22$  & 0.42 \\   [0.3em]
$F(h) \propto \sin(\pi h/2h_\mathrm{out})$ & 1     & $1/2$ & 1.57 & $2/9$ & $ 0.54$ \\  [0.3em]
$t_\mathrm{vis}$ independent of  $r$ & 2      & $\infty$ &  ---  & ---  & --- \\
\hline    
\end{tabular}
\end{table*}
  For a specific initial distribution of the viscous torque $F_0(r) = 3\pi h \nu \Sigma_0(r)$, the accretion rate can be found as 
 \begin{equation}
\dot M(h,t) = \int\limits_{h_{\rm in}  }^{h_{\rm out}  } \,G_{\dot M}(h,h_1, t) \, F_0(h_1)\, \mathrm{d}
h_1/h_{\rm out}\, ,
\label{eq:Mdot_from_green}
\end{equation}
where $h=\sqrt{GMr}$ and  $G_{\dot M}$ is calculated as a convergent series. For large time $t$, one term dominates in $G_{\dot M}$ and the time dependence can be written as one exponential
$$
\dot M (0,t) \propto G_{\dot M}(0,h_1,t)\Big|_{t> t_\mathrm{vis}} 
 \propto ~
 \exp\left(-  \frac{t }{t_\mathrm{exp} }\right)\, ,
$$
\begin{equation}
t_\mathrm{exp}  = 
\frac{16 \, l^2}{3\, k_1^2}\,
\frac{r_\mathrm{out}^{2}}{{\nu_\mathrm{out}}}~,
\label{eq:texp}
\end{equation}
where $\nu_\mathrm{out} = \nu_0 \, r_\mathrm{out}^b$ and $2\,l\,(2-b)=1$. 

In Table~\ref{tab:zeroes} some parameters of solutions with $m=0$ are presented, when the
outer boundary condition $\dot M(\rout) = {\partial F}/{\partial h}= 0$ and the inner boundary condition $F(\rin)=0$.
The peak time for a bounded disc is of the order of $t_{\rm max}$ for a freely expanding disc, but it also depends on the outer radius $r_{\rm out}$~\citep{lipunova2015}. The disc becomes quasi-stationary (namely, the accretion rate virtually does not change with radius) in the region limited by $r< r_\mathrm{out}  \times (t/t_\mathrm{exp})^{2\, l}$. The establishment of quasi stationarity in the central regions of  discs on viscous time scales is a common property for discs with any type of viscosity.

\citet{Nixon-Pringle2021} found the Green's function for $\nu \propto r$ and a general inner boundary condition~\eqref{eq.general_inner_cond}. They assume the outer boundary condition $F(\rout) = 0$ or $\Sigma(\rout) = 0$, so that the mass and angular momentum reaching the outer
boundary is absorbed there. 

Some bright outbursts in X-ray novae show lightcurves with fast rise and exponential decay~\citep[FRED lightcurves,][]{chen_et1997}. \citet{kin-rit1998} studied the evolution of a disc with the finite outer radius and constant $\nut$, and found that the accretion rate declined exponentially with time. 
This does not necessarily mean that the viscosity is constant in real discs. The FRED light curves are equally well approximated within the model of evolving  $\alpha$-discs with a time-dependent viscosity $\nu$. 
This is due to the fact that (i)  on time scales of the order of one to two $t_\mathrm{vis}$, an $\alpha$-disc and a disc with time-independent viscosity show similar evolution (compare the solid curves in Fig.~\ref{fig:analyt_colutions_Mdot}); (ii) the typical spectral evolution of an X-ray Nova in the soft state implies a softening of the spectrum, characterized then by an exponential tail. Consequently, a power-law variation of the disc maximum temperature is masked by the exponentially decreasing flux in the observational band~\citep{lipunova_shakura2000}. In any case, before applying an analytic model to an outburst of an X-ray nova, it is necessary to ascertain whether the radius of the high-viscosity zone in the disc is constant or the cooling front is propagating inwards,
see \S\ref{sec4}.

It is possible to approximate the evolution of an $\alpha$-disc with $\rout=\,$const using an exponential solution for time-independent viscosity \eqref{eq:texp}. For this,  it is 
sufficient to estimate the most appropriate value of the parameter $b$.
 This can be done using the relation  between the kinematic viscosity and $\alpha$, which follows from \eqref{eq:alpha_prescription}:
 \begin{equation}
\nu = \nu_0 \, r^{b} \sim \frac 23\, \alpha\, \omegak\, r^2\, 
\left(\frac{z}{r}\right)^2\, .
\label{eq.nut_alpha_z_r}
 \end{equation}
Stationary $\alpha$-discs which are gas-pressure dominated and have Kramers opacity have an aspect ratio $z/r \propto r^{1/8} \dot M^{3/20}$ \citep[][]{sha73}. 
Thus $b= 3/4$ if we ignore the dependence of the disc thickness on the accretion rate.  Further options for $\alpha$-discs are listed in Table~\ref{tab:zeroes}.
Potentially, knowing the observed time $t_{\exp}$, one can estimate $\alpha$ in an X-ray nova using~\eqref{eq:texp} and \eqref{eq.nut_alpha_z_r}~\citep{lipunova2015, lipunova-malanchev2017}. 
However, the main uncertainty arises from the unknown outer radius of the high-viscosity zone. The most promising approach to determine $\alpha$ observationally lies in examining the FRED outbursts of the short-period BH XRBs, where there is a high chance that $r_{\rm out}$
coincides with the outer radius of a tidally truncated disc.

\subsection{Solutions to nonlinear evolution equation for freely expanding discs}
 \label{s.non-lin-equation-freely}

 \index{viscous disc evolution}
Earlier we considered a scenario in which the kinematic viscosity coefficient depends only on the radial coordinate inside the disc. For many viscosity mechanisms, we can express $\nu$ as a function of a power law of $\Sigma $ and $r$.
Consequently, the viscous time in the disc varies with radius and time. This is relevant, for example, for discs with $\alpha$-viscosity.
For solving the nonlinear differential equation of disc evolution~\eqref{eq.nonlin-diff}, similarity methods have proven to be quite effective.

Self-similar solutions to nonlinear differential equations can be divided into two kinds~\citep{barenblatt1982e,barenblatt2003scaling}. The self-similar solution is of the first kind if the self-similar function, as well as its dimensionless argument, can be derived from dimensional analysis or conservation laws. 
This case is also called a complete self-similarity.
This is, for example, J. I. Taylor's blast wave problem with energy conservation. 

The second kind, or incomplete self-similarity, arises when it is not possible to use dimensional analysis to determine the self-similar function, or to find the powers to which the dimensional parameters should be raised to produce a self-similar dimensionless variable.  In this case, the self-similar function is found as a particular solution to the problem itself~\citep[a `nonlinear eigenvalue problem',][]{zeldovich-raizer1967}. 
The self-similar solution describes the `intermediate asymptotic' behavior of the system in the region where it no longer depends on the initial and boundary conditions.

Self-similar solutions of the first kind have been found by~\citet{pringle1974PhD,pringle1991} for the nonlinear viscous diffusion equation of accretion discs in the evolutionary stage when the accretion rate decays.
The conservation of total angular momentum in an accretion disc is used there.

Solutions of the second kind have been obtained by~\citet{lyub-shak1987} for the earlier evolutionary stage when the central accretion rate increases.

\begin{figure}
    \centering
\includegraphics[width=0.6\linewidth]{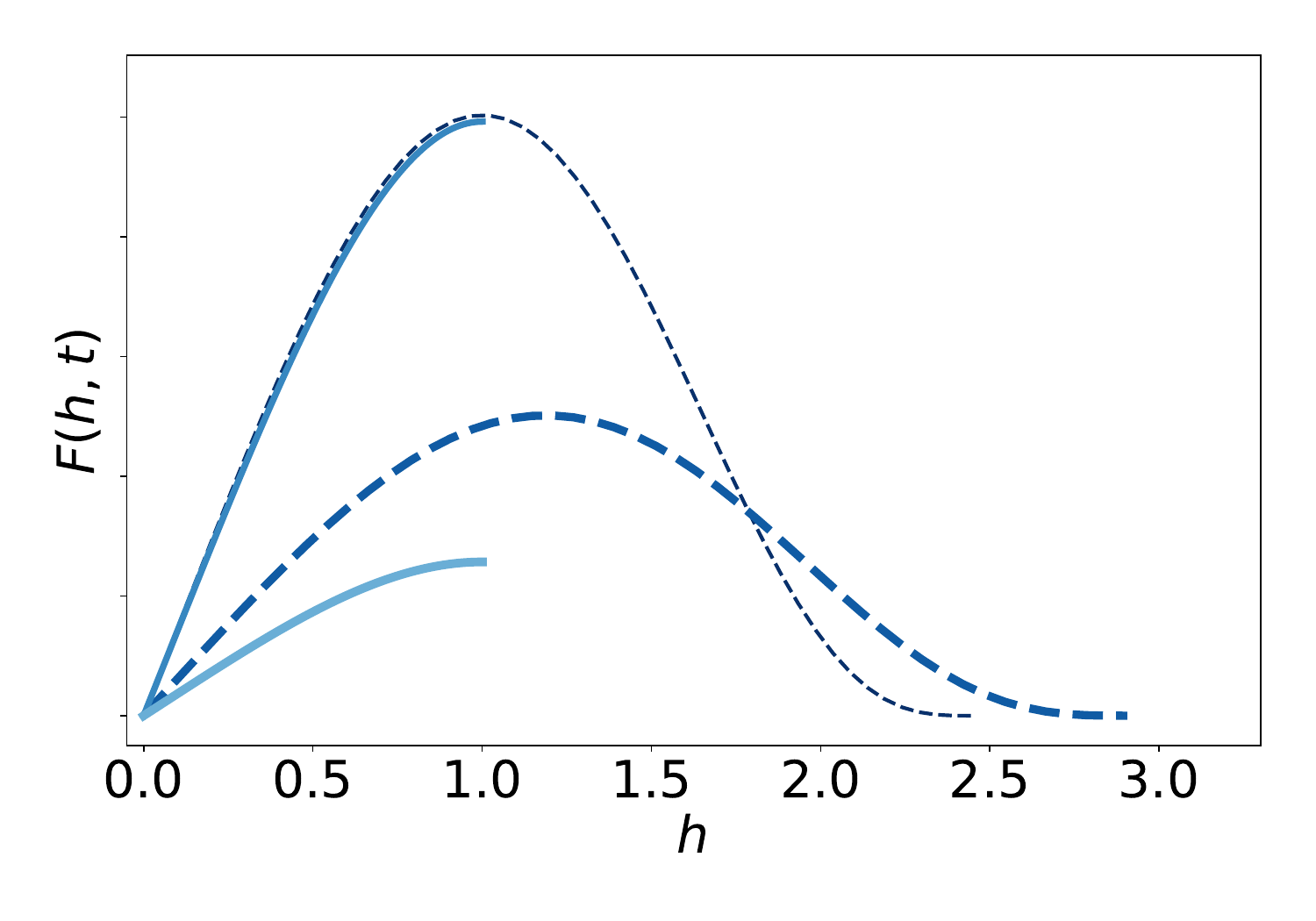}
    \caption{Distributions of viscous torque $F$ versus angular momentum $h$ at two moments of time in $\alpha$-discs, freely-expanding  (dashed) and limited by the outer radius with $h_{\rm out} = 1$ (solid). The accretion rate on the center decays with time.
}
\label{fig:shape_alpha_disc}
\end{figure}

\citet{lyub-shak1987} proposed a division of the whole evolution of a ring into three stages, each of which allows self-similar solutions: (1) the stage of initial spreading of the disc from a ring, (2) the gradual development of a quasi-stationary distribution of parameters in the disc, and (3) the late `spreading' of the disc away from the center, accompanied by a decrease in the central accretion rate (see the dashed lines in Fig.~\ref{fig:shape_alpha_disc}).

In the first and second stages, intermediate asymptotics have been found which are valid in the regions of the disc far from the boundaries (the initial ring position and the last marginally stable orbit around a black hole or the magnetospheric boundary). 
The self-similarity index is found not by dimensional arguments, but by integrating the ordinary differential equation for the representative function. 
Meanwhile, in the outer region, the conditions remain close to the original ones, because the viscous times are longer at larger radii. During the 1st stage, when the central accretion rate is still zero, the asymptotic for the inner disc radius is found. During the 2nd stage, a power-law dependence of the increasing accretion rate is obtained. 

The disc gradually evolves into the third and final stage (the {decay} stage), when the details of the initial distribution are `forgotten' anyway, and only some integral quantities conserved by the evolution are important in finding the self-similar solution.\footnote{The nonlinear problem  has the following distinctive features. Firstly, the self-similar solutions of the second kind exist only for $m\neq 0 $. 
Secondly, self-similar solutions of the first kind in the third stage, while they exist for $m=0$, have an exponential radial profile for $r\rightarrow \infty$, characteristic for linear problems~\citep[see, for example,][and Fig.~\ref{fig:green_functions}, the left panel]{lyn-pri1974}.
For $m\neq 0 $, the position of the outer boundary of the disc is fully determined, 
see the dashed profile in Fig.~\ref{fig:shape_alpha_disc},  also \citet{rafikov2016}. This property is similar to the one that arises in problems of thermal conductivity, when, due to the nonlinearity, the heatwave boundary sharply separates the heated zone from the rest of the region~\citep{zeldovich-raizer1967}.}
The solution is sought in the form 
$F={h^{({n+2})/{m}}}\,{(\pm D\, t)^{-1/m}} \, y(\xi)$ where $y$ is a dimensionless function of a dimensionless argument $\xi = {h}/({A\,(\pm t)^{\beta}})$. The index $\beta$ can be determined in the course of the solution for the first and second stage, and from the conservation of the angular momentum, for the decay stage (then $\beta$ can be expressed algebraically, see also \citealt{pringle1974PhD}).
The accretion rate on the gravitating center decays as follows~\citep[see., e.g., chapter 1 in][]{shakura_etal2018}\footnote{For $a=0$ and $m=0$, Eq.~\eqref{eq:Mdot_law_unbounded} transforms into the \citeauthor{lyn-pri1974} solution \eqref{eq.LBP_solution_Mdot} at large $t$.} 
:
\begin{equation}
\dot M(t)  = {\dot M}_0\,  (1+t/\tilde{t_0})^{-1-\beta}\, ,
\qquad 
\beta = \cfrac{1-m}{n+2} = \left( 5\,a-2\,b+4 \right) ^{-1}\, ,
\label{eq:Mdot_law_unbounded}
\end{equation}
where $\dot M_0$ is the central accretion rate at $t=0$. Time zero can be assigned to any point during the decay stage. 
It is possible to relate the normalization time $\tilde{t_0}$ with some viscous time\footnote{For a nonlinear problem, the viscous time depends on both the radius and time.}. One can consider the viscous time ${r_\mathrm{c}^{2}}/{{\nu_\mathrm{c}}} $ at moment $t=0$ at the radius $r_\mathrm{c}$, where the asymmetric bell-shaped distribution of the viscous torque has a maximum\footnote{This is the boundary between accretion and decretion, where $\partial F/\partial h=0$.},  and see that 
\begin{equation}
\tilde{t_0}= \frac{4}{3} \, \beta\,
\frac{r_\mathrm{c}^{2}}{{\nu_\mathrm{c}}}\, .
\label{eq.t_0_tilde} 
\end{equation}

\begin{table}
 \centering
 \caption[Dimensionless parameters in the equations of non-stationary accretion]{Parameters of the time-dependent solutions for discs with the zero inner torque  and viscosity $\nu \propto \Sigma^a \,r^b $, $a\neq 0$. A freely expanding disc evolves according to $\dot M \propto  (1+t/\tilde{t}_0)^{-1-\beta}$, see \eqref{eq:Mdot_law_unbounded} and \eqref{eq.t_0_tilde}.  In the disc with the fixed outer radius the accretion rate evolves as $\dot M\propto (1+t/t_0)^{-1/m}$, see~\eqref{eq.t_0}, unless $P_{\rm tot} = P_{\rm rad}$.  }
 \label{tab.indexes}

\begin{tabular}{@{}p{4.5cm}cccccc@{}}
\hline
~~~~~~~~~~Disc type & $a$ & $b$ & $1+\beta$ & $m$ & $n$ &$\lambda$ \\
\hline
$\alpha-$disc, $\kappa =$ const, $ P_{\rm tot} = P_{\rm gas}$ & $2/3$ & 1 & $19/16$ & $2/5$ & $6/5$ & $3.482$\\[0.3em]
$\alpha-$disc, $\kappa =$ const, $ P_{\rm tot} = P_{\rm rad}$ & -2 & 3/2 & (-8/9) & $2$ & $7$ &  --- \\[0.3em]
$\alpha-$disc, $\kappa \propto \rho T^{-7/2}$, $ P_{\rm tot} = P_{\rm gas}$ & $3/7$ & $15/14$ & $5/4$ & $3/10$ & $4/5$ & $3.137$ \\[0.3em]
$\alpha-$disc, $\kappa \propto \rho T^{-5/2}$, $ P_{\rm tot} = P_{\rm gas}$ & $1/2$ & $1$ & $11/9$ & $1/3$ & $1$ &$3.319$\\[0.3em]
\makecell{convective turbulence, $\kappa \propto T^2$,~~~~~~\\ \citet{lin-bodenheimer1982}} & 2 & 0 & $15/14$ & $2/3$ & $8/3$ & 4.820\\[0.3em]
\makecell{molecular disc with gravitational~~~~\\
instability,~\citet{lin-pringle1987}} & 2 & $9/2$ & $6/5$ & $2/3$ & $-1/3$ &1.788\\[0.3em]
\hline    
\end{tabular}
\end{table}

In the gas-pressure dominated $\alpha$-disc with the opacity according to the Kramers law and Thomson scattering, the accretion rate decreases as $\propto t^{-19/16}$ and $\propto t^{-5/4}$, respectively~\citep{pringle1974PhD,filipov1984,lyub-shak1987,cannizzo_etal1990,pringle1991}.
Actually, an opacity $\kappa \propto \rho T^{-5/2}$ better describes $\alpha$-discs around stellar-mass objects for conditions  near the outer radius of the hot ionized zone~\citep{tavleev+2023}.
Other solutions in Table~\ref{tab.indexes} are important in the context of protoplanetary discs and discs in galactic nuclei.
\citet{lin-pringle1987} considered a molecular disc with a gravitational instability producing an effective viscosity. 
A different approach to describe analytically viscosity in self-gravitating discs, depending on their thermal structure and external irradiation, can be found in \citet{Rafikov2009, Rafikov2015}. 
\citet{lin-bodenheimer1982} studied the evolution of a protoplanetary disc under the influence of convective turbulent viscosity.

\citet{pringle1991}  considered an ``external'' disc with a zero central accretion rate and a finite central torque, w~\citep[a massive binary surrounded by a gas disc, see also][]{ivanov_etal1999}. For all mass initially at the origin and being gradually expelled to infinity, the inner torque decays as 
$$
F(0) \propto {t}^{ -1 + l}\ , \quad 
l =\left( 4\,a-2\,b +4\right) ^{-1}\, .
$$
The latter dependence converges to the solution for the linear equation with $a=0$, see \S\ref{s.linear_freeexp}.
In~\citet{rafikov2013}, self-similar solutions were found with a non-zero constant mass  accretion rate at the inner boundary $\dot M(R_{\rm in}) = \chi \dot M_\infty $  and a constant external mass supply $\dot M_\infty $. Larger $\chi$ values imply less mass accumulation in the disc and a smaller inner torque. 
General late-time power-law asymptotics for freely expanding discs with simultaneously non-zero central accretion rate and non-zero central viscous torque at the inner boundary were presented by \citet{rafikov2016}.
Formulating such solutions required setting the inner coordinate of the disc to zero.

A time-dependent solution of \citet{ogilvie1999} for accretion flows that retain the heat generated by viscous dissipation --- 'advection-dominated' flows -- is another example of using similarity methods.   In the case of quasi-spherical accretion, the governing equations (which are spherically averaged) are nonlinear, so that the self-similar solution of the first kind is an attractor for the initial-value problem with conserved angular momentum.  The accretion rate varies $\propto t^{-4/3}$, which is consistent with the solution of the thin-disc evolution equation with $z/r = $const, see Table~\ref{tab:zeroes}.

 \subsection{Non-linear evolution equation for  $\rout=\,$const}
 \label{s.non-lin-equation-bounded}

\newcommand{\appropto}{\mathrel{\vcenter{
  \offinterlineskip\halign{\hfil$##$\cr
    \propto\cr\noalign{\kern2pt}\sim\cr\noalign{\kern-2pt}}}}}
\begin{figure}
    \centering
    \includegraphics[width=0.85\linewidth]{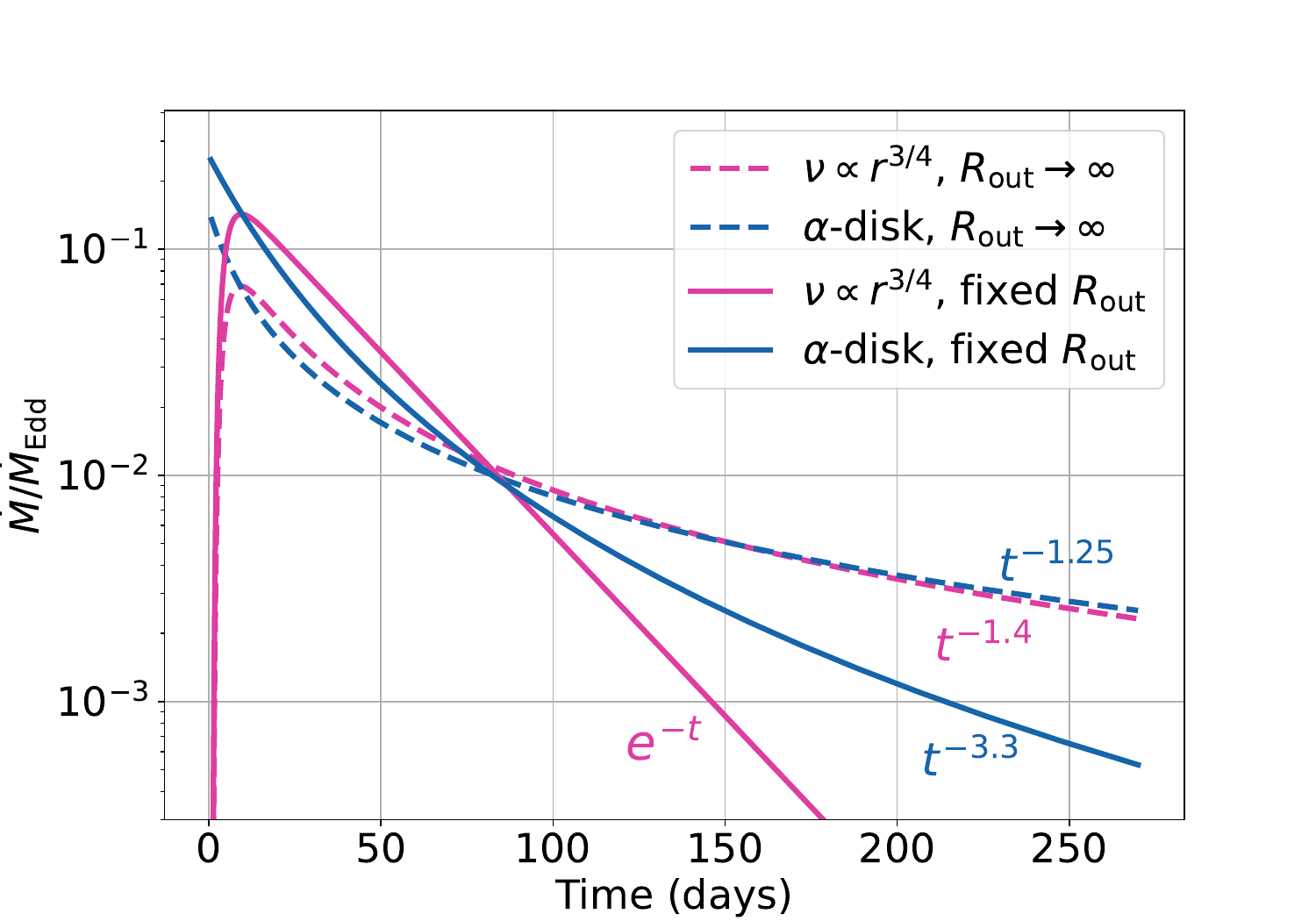}
    \caption{Comparison of analytic solutions \eqref{eq:LBP_mdot}, \eqref{eq:texp}, \eqref{eq:Mdot_law_unbounded}, and  \eqref{eq:Mdot_fixedRout_alpha} for the evolution of the accretion rate onto a 10~$M_\odot$ black hole with different viscosity  and disc outer boundary (freely expanding or fixed). The viscosity in each solution is chosen to approximate the law in the zone of dominant gas pressure and opacity according to the Kramers formula. At late times, the time-dependencies tend to $ (t/\tau_\mathrm{pl})^{-7/5}$, $\exp(-t/t_\mathrm{exp})$, $ (1+t/\tilde{t}_0)^{-5/4}$, and  $(1+t/t_0)^{-10/3}$, respectively. The input parameters are  $M_{\rm disc} = 6\times 10^{24}~$g, $R_{\mathrm{out}}=2\times 10^{11}$~cm and $t_{\rm vis} = 60^{\mathrm d}$. The corresponding $e$-decay time  is $t_\mathrm{exp}\approx 27^{\mathrm d}$; the bounded $\alpha$-disc's solution is close to the linear-problem solution for about $3\, t_\mathrm{exp}$.
}
\label{fig:analyt_colutions_Mdot}
\end{figure}
 A solution to the basic equation of non-stationary accretion \eqref{eq.nonlin-diff} for a disc with a constant outer radius can be found during the decay stage using the method of separation of variables: $F(h,t)=F(t)\times f(h/\ho)$.
The accretion rate changes proportionally to $F(t)$ \citep{Ludwig+1994, lipunova_shakura2000, Ritter-King2001}: 
\begin{equation}
\dot M (t) = \dot M_0 \, (1+t/t_0)^{-1/m}\, ,
\label{eq:Mdot_fixedRout_alpha} 
\end{equation}
where $\dot M_0$ is the central accretion rate at $t=0$, which can be chosen as any time at the decay stage. 
 The space part of the solution $f(h/\ho)$ for a disc with zero (or very small) viscous stress at the inner boundary was found semi-analytically as a polynomial
 by \citet{lipunova_shakura2000} (see Fig.~\ref{fig:shape_alpha_disc}, the solid curves). The outer boundary condition was set to $\dot M(\rout) = 0$, which is relevant during outbursts of X-ray novae, when the accretion rate inside the disc is much higher than the mass transfer rate from the companion star. The full solution allows determination of  $t_0$, which is of the order of the viscous time at the disc outer radius:
\begin{equation}
t_0 = \frac{ h_{\rm out}^{n+2-m}\, a_0^m}{\lambda\,m\, D\, \dot M^m_0} = \frac{4}{3\lambda\,a} \,
\frac{r_\mathrm{out}^{2}}{{\nu_\mathrm{out}(t=0)}}\, 
\label{eq.t_0} 
\end{equation}
and the time scale on which the disc mass decays initially~\citep{Ritter-King2001}.
Here $h_{\rm out} = \sqrt{GMr_{\rm out}}$, $a_0\equiv f'(0)$, $a = m/(1-m)$, and 
$\lambda\sim 3$ (see Table~\ref{tab.indexes}).
The solution for Kramers opacity was used to model the optical and X-ray lightcurves of the X-ray novae \hbox{A\,0620-00} and \hbox{GU~Mus\,1124-68} during the decline after the peak of their outbursts and to obtain constraints on the turbulence parameter $\alpha$~\citep{lip-sha2002,suleimanov_etal2008}.

Characteristically, a disc with a fixed outer radius evolves faster than a freely expanding disc with the same viscosity (compare the curves of the same color in Fig.~\ref{fig:analyt_colutions_Mdot}). This is understandable as the time scale in a freely expanding disc --- the longest viscous time --- continues to increase with time just as the size of the disc does.

\subsection{A note on analytic solutions}

 The analytic solutions are applicable under the assumption  
 that viscosity is a power law function of radius and surface density.
 These models can serve as benchmarks for testing the behavior of numerical schemes. However, in real astrophysical discs, a region where the viscosity can be considered to be smoothly varying 
 is in most cases not only limited, but also changes in size on a time scale comparable to the viscous time scale.  
 In accretion discs surrounding stellar-mass compact objects, viscosity due to MRI-driven turbulence is believed to be suppressed in cooler recombined regions away from the gravitational center. Cold, massive discs --- such as protoplanetary discs or those found in active galactic nuclei --- are often characterized as multizone structures,
 where the viscosity due to gravitoturbulence depends on the details of thermal balance and external irradiation.
 
 When studying disc dynamics during outbursts in binary systems, where discs are tidally truncated, solutions obtained with $\rout=\,$const are expected to be more relevant. Even in these cases, the changing temperature may become insufficient for full ionization in the outer disc regions. \footnote{$\rout =\,$ const may hold temporarily for some outbursts, very bright ones, or in very short-period binary systems.} Furthermore, recombination leads to thermal instability and subsequent changes in the disc structure on timescales shorter than the viscous one.  For example, as the accretion rate decays during an X-ray or dwarf nova outburst, it is inevitable that the region of the hot, ionized disc will progressively contract. This phenomenon leaves a distinct imprint on the light curve shapes and emphasizes the necessity for numerical simulations of time-dependent discs, such as those implemented in the disc instability model (DIM) code~\citep[for review, see][and also \S\ref{sec:dndim}]{lasota2001,Hameury-Lasota2020} and \texttt{Freddi} code~\citep[available on \texttt{github}]{lipunova-malanchev2017}.

\section{Angular Momentum Transport by Magnetic Fields}\label{sec3}

Accretion can only happen in a rotationally-supported disc if there is some mechanism for extracting angular momentum from the rotating fluid elements.  This is why the Shakura-Sunyaev ansatz (\ref{eq:alpha_prescription}) has been so widely used in building analytic models of accretion discs.  Over the last fifty years, significant progress has been made in understanding the basic physics of this angular momentum-transporting stress, and the advent of large-scale numerical simulations has created the opportunity for building models that incorporate this basic physics.

While nonaxisymmetric spiral waves can contribute to hydrodynamically transporting angular momentum outwards (see, e.g., section 5 of \citet{pap95} for a detailed review), it appears that in discs with high microscopic electrical conductivity and high orbital Mach numbers, magnetic stresses will dominate \citep{ju16,ju17}.  There are three mechanisms whereby magnetic fields can transport angular momentum, which are not mutually exclusive:  magnetorotational (MRI) turbulence, mean-field magnetic stresses within the disc, and magnetocentrifugal winds.\footnote{There is a semantic issue here in that it can be argued that any form of magnetic angular momentum transport is magnetorotational in nature.  We are here drawing a distinction between local turbulence and more global mean field transport.  It is only the former that {\it might} be describable in terms of the local Shakura-Sunyaev prescription (\ref{eq:alpha_prescription}).}

\subsection{Magnetorotational (MRI) Turbulence}
\label{sec:MRI}

A major breakthrough in accretion disc theory occurred in the early 1990's with the realization that the magnetorotational instability  \citep[MRI;][]{bal91,haw91,haw92,bal92} provides a robust mechanism for exciting turbulence in a disc with precisely the necessary properties to both transport angular momentum outward between fluid elements and to dissipate mechanical energy into internal energy \citep{bal99}.  An excellent review of the physics of this instability can be found in \citet{bal98}.  The simplest version consists of axisymmetric perturbations on a vertical magnetic field, a situation that is always unstable in the presence of negative radial angular velocity gradients provided that the field is weak enough that the vertical Alfv\'en wave period is longer than of order the orbital period.  Because the wave period is proportional to the vertical wavelength, we require that the minimum unstable mode wavelength $\sim v_{{\rm A}z}/\Omega$ can fit within some measure of the size of the disc, particularly the disc scale height.  Here $v_{{\rm A}z}$ is the local vertical Alfv\'en speed and $\Omega$ is the local angular velocity.  If the disc is supported vertically by thermal pressure, then this requires the magnetic pressure due to the vertical magnetic field to be subthermal, i.e. plasma betas greater than unity.  The growth rate can be greatly reduced by curvature of a strong toroidal field when the toroidal Alv\'en speed exceeds of order the geometric mean of the sound and orbital speeds \citep{pes05,das18}. However, even a purely toroidal magnetic field will be unstable to transient growth of nonaxisymmetric perturbations in the presence of negative radial angular velocity gradients \citep{bal92,fog95,ter96}.\footnote{Exponential growth of true nonaxisymmetric unstable modes can only exist in a rotating shear flow when one solves the global eigenvalue problem with boundary conditions.  Unfortunately, linear analyses (axisymmetric or nonaxisymmetric) beyond the simplest local instability can be complicated, particularly when they account for the vertical or global structure of the disc, and/or the coupling to other instabilities \citep{gam94,fog95,ter96,cur96,das18,kim00}.}

\begin{figure}
    \centering
    \includegraphics[width=0.9\linewidth]{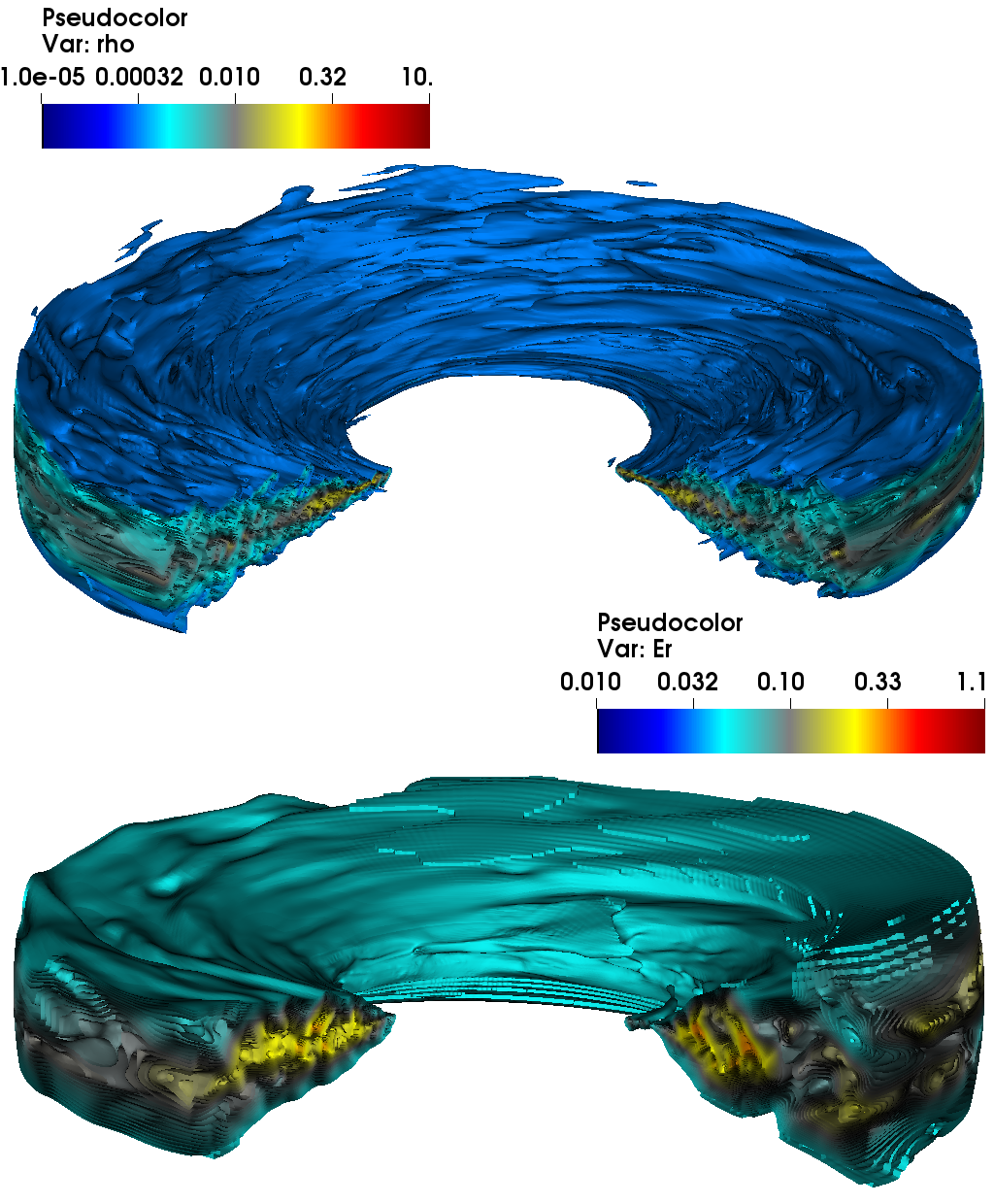}
    \caption{Snapshot of the three dimensional structure of mass density (top) and radiation energy density (bottom) of a global simulation of MRI turbulence in a disk around a supemassive black hole \citep{jia20}.  In addition to MHD, this simulation incorporates angle-dependent radiation transport with frequency-averaged (grey) opacities.}
    \label{fig:AGNIron}
\end{figure}

Over the years, substantial effort by many groups has been made to understand the nonlinear outcome of the MRI in three computational domains:  local shearing box simulations of a rotating patch of the disc \citep{haw95} that do not include (vertically unstratified) or do include (vertically stratified) vertical gravity, and global simulations (see Fig.~\ref{fig:AGNIron} for an example).  These simulations start with a laminar flow with some initial magnetic field that then becomes unstable and turbulent. The vertically stratified and global simulations of weak field MRI exhibit dynamo behavior in which the radial and toroidal fields alternate polarity and rise with height in the disc \citep{bra95} --- the so-called MRI butterfly diagram.   An excellent modern review of the dynamo properties of MRI turbulence can be found in section 5.3 of \citet{rin19}.  Nevertheless, the topology of the initial magnetic field tends to have a long term effect on the properties of the turbulence and the turbulent stresses.  For example, in shearing box simulations, the saturated turbulent stress increases above a certain level with increasing net vertical magnetic flux until the MRI is stabilized \citep{haw95,pes07}.  In global simulations the long term outcome tends to depend on how much local vertical flux is present.  This dependence on initial field topology is unfortunate given that we have little information concerning the magnetic field geometry in the material that fuels accretion discs across astrophysics.  In recent work, however, vertical flux was self-consistently generated by a dynamo operating on an initially purely toroidal field configuration \citep{jac24}.

The resistive and viscous dissipation scales in the turbulent cascade are often impossible to resolve with existing computational resources, and even getting enough dynamic range to produce a convincing inertial range has only recently been achieved \citep{kaw24}.  Hence the vast majority of simulations are run in such a manner as to have magnetic and kinetic energy dissipate at the grid scale - so-called ILES (Implicit Large Eddy Simulations) methods.  Unfortunately, this is also not without consequences.  Zero net flux, vertically unstratified shearing box ILES simulations with simplified (isothermal) thermodynamics of MRI turbulence do not converge:  the saturation level of the turbulent stresses decreases with increasing numerical resolution \citep{pes07,fro07a}.  This also appears to be the case when vertical stratification is included \citep{rya17}.  However, when explicit viscosity ($\nu$) and resistivity ($\eta$) (with values much larger than are realistic) are included, then numerical convergence is obtained with zero net magnetic flux \citep{fro07b,les07}.  The resulting stresses exhibit a strong dependence on magnetic Prandtl number $\mathrm{Pm}\equiv\nu/\eta$.  This dependence can even result in thermal/viscous instabilities with stable upper and lower temperature branches.  This has been explored in the context of the inner regions of black hole X-ray binary discs by \citet{pot17}.  A compelling explanation for the Prandtl number dependence in zero net flux, unstratified shearing boxes has been recently presented by \citet{mam20}.  This has also been extended to vertically stratified shearing boxes at high magnetic Prandtl number (the regime of relevance to luminous active galactic nuclei and the inner regions of X-ray binaries) by \citet{hel22b,hel24}.

Provided explicit dissipation coefficients are included and the box size is sufficiently large, unstratified zero net flux simulations even produce saturated turbulent stresses that are proportional to the thermal pressure \citep{ros16}, in agreement with the alpha prescription.  This dependence is weakened, however, in the presence of net magnetic flux through the box.

Thermodynamics can also affect the properties of MRI turbulence.  In a radiation pressure dominated environment, such as in the innermost region of a standard Shakura-Sunyaev disc, photon diffusion can cause the turbulence to be highly compressible even though the turbulent fluid speeds are much less than the total sound speed in the radiation-dominated plasma \citep{tur03}.  This is because photon diffusion out of a compressing region reduces the build-up of radiation pressure to resist the compression.  This diffusion is also accompanied by significant damping of compressible motions \citep{ago98}, i.e. a large radiation bulk viscosity.  As a result, the turbulence has a high effective magnetic Prandtl number, and the ratio of Maxwell to turbulent Reynolds stress is enhanced \citep{jia13a}.

In addition to mediating the dissipation of accretion power through a turbulent cascade, MRI turbulence can also affect the cooling rate of an accretion disc in the radiation pressure dominated regime.  The large density variations in the compressible turbulence create porosity that enhances the photon diffusion rate, with photons diffusing mostly through low density channels.  (This effect has been quantified in compressible convective turbulence in massive stars by \citet{sch20}.)  In addition, buoyant magnetic field concentrations that form in the turbulence can produce bulk transport of photons through vertical advection \citep{bla11,jia14}.

Vertically stratified shearing box simulations of MRI turbulence that incorporate radiation transfer with temperature-dependent opacities can become vertically unstable to convection when the opacities are high enough \citep{hir14,hir15,col17,sce18a}.  Depending on the behavior of the opacity with temperature, the simulations can alternate between convective and radiative cycles or maintain persistent convection.  When convection is present, the alternating polarity of azimuthal field so characteristic of the MRI turbulent dynamo in vertically stratified shearing boxes \citep{bra95} is suppressed, likely by vertical mixing of field \citep{col17}.  In some \citep[but not all,][]{sce18a} cases, this convection can enhance the time-averaged MRI turbulent stresses even in the absence of net vertical magnetic flux.  All of this behavior has also been observed in global radiation MHD simulations of MRI turbulence with temperature-dependent opacities \citep{jia20}.  The enhancement of stresses appears to be associated with the creation of vertical magnetic field by convection \citep{hir14}, but more research would be helpful to fully understand this.  Recent numerical experiments using non-radiative shearing box simulations but with fixed resistivity and fixed optically thin cooling time in the uppermost layers (to control the vertical entropy gradient) find that the flow can alternate between convection and MRI-dominated episodes, with an enhancement of stress during the MRI phases (with little or no convection at those times) compared to simulations in which convection never takes place \citep{hel21}.
Note that in all these cases, the angular momentum transport appears to still be fundamentally magnetic in nature.  Non-magnetic hydrodynamic simulations of convection in disks tend to give rise to much smaller transport, with $\alpha$ ranging from $10^{-6}$ to $10^{-5}$ \citep{hel18}.  Moreover, early simulations actually
found the transport to be inward, not outward \citep{cab96,sto96}.  Only at sufficiently high Rayleigh numbers does direct angular momentum transport by hydrodynamic convection become outward \citep{les10,hel18}.

Although MRI is generally thought to be a weak field instability, simulations with substantial vertical flux can lead to configurations with strong turbulence and in which toroidal magnetic pressure dominates thermal pressure and provides hydrostatic support of the disc against vertical gravity \citep{bai13,sal16a,sal16b}.   Whether this is still the same as standard MRI turbulence is unclear: the regular toroidal field reversals so characteristic of the weak field MRI dynamo (the butterfly diagram) are lost in this regime.  The vertical structure of such discs appears to be consistent with some of the analytic ideas proposed for magnetically dominated discs \citep{par03,beg07,beg15}, and it is likely that such discs are thermally and viscously stable.

Strongly magnetized, turbulent structures can also arise in simulations that start with no vertical magnetic field, provided a sufficiently strong toroidal field is present.  This was perhaps first anticipated by \citet{tou92}, who proposed a dynamo mechanism coupling the Parker instability, MRI, and magnetic reconnection.  That such a thing is possible was first demonstrated numerically by \citet{joh08}: a strong toroidal field that vertically supports the disc is Parker unstable, creating vertical magnetic field that is MRI unstable.  The nonaxisymmetric MRI on the toroidal field (which does {\it not} require a weak field) also plays a role in the resulting nonlinear state.  Conducting boundary conditions that preclude vertical field escape were used for these simulations \citep{joh08}, and simulations with outflowing boundary conditions, both local \citep{sal16b} and global \citep{fra17} were unable to retain strong toroidal fields and replicate this result.  On the other hand, various forms of outflow boundary conditions were implemented in recent vertically stratified shearing box simulations with pure toroidal fields \citep{squ24}, and these succeeded in producing self-sustaining strongly magnetized turbulent configurations, like those in the original work of \citet{joh08}. Without vertical field, the midplane must be magnetically dominated in the saturated state for this to happen, otherwise buoyant field loss returns the system to a weakly magnetized state of MRI turbulence.  Much more work needs to be done to understand the origin of this strongly magnetized regime and how it might play out in different astrophysical environments.

\subsection{Mean-Field Magnetic Stresses Within the Disc}
\label{sec:meanfieldstresses}

Depending on initial conditions, global simulations of magnetized accretion flows can lead to states where magnetic pressure dominates thermal pressure.   These tend to have large scale, coherent magnetic fields \citep{mac06,gab12,hop24b}, in which mean Maxwell stresses are comparable to or even dominate the outward angular momentum transport compared to turbulent Maxwell stresses, i.e. $\langle B_rB_\phi\rangle\sim\langle B_r\rangle\langle B_\phi\rangle$.  Perhaps the most extreme version of this regime has come from recent zoom-in cosmological simulations of the fueling of a supermassive black hole \citep{hop24a}.  Midplane plasma beta's between $10^{-6}$ and $10^{-2}$ exist in the disc on a scale of $\sim100$ gravitational radii.  The field is predominately toroidal and is maintained by inward advection of magnetic flux.  The Alfv\'en speed greatly exceeds the geometric mean of the sound and orbital speeds, which as we discussed above is the limit where MRI growth rates are significantly reduced \citep{pes05,das18}.  Nevertheless, strong turbulence exists in the disc, which is supported vertically by both magnetic pressure and turbulent kinetic energy gradients \citep{hop24b}.  Analytic scalings of these simulation results can be found in \citet{hop24c}.

A hybrid situation has also been found in simulations with net poloidal flux in which most of the accretion occurs in a magnetically dominated ``corona'' that sandwiches an equatorial, thermal pressure-dominated, MRI turbulent disc, and where angular momentum transport in the corona is dominated by mean field Maxwell stresses \citep{suz14,zhu18,mis20}.  This high altitude flow can even consist of alternating layers of turbulent and mean field Maxwell stresses, with the turbulent layers being due to MRI \citep{jac21}. 

\subsection{Magnetocentrifugal Winds}

A number of authors in the 1970's \citep{bla76,lov76} proposed that double-lobed radio sources could be the result of energy and angular momentum losses in a magnetohydrodynamic wind emerging from the upper and lower faces of an accretion disc.  The dynamical structure of this wind was fully developed in the force-free limit by \citet{bla76} and in magnetohydrodynamics by \citet{bla82}, with the disc as a lower boundary condition.  Accretion via angular momentum losses in a magnetocentrifugal wind has received considerable attention in the context of weakly ionized flows, such as protoplanetary discs, where the interior of the disc may not be able to sustain MRI turbulence \citep[see, e.g.,][ and references therein]{les21}.

Vertically stratified shearing box simulations with relatively strong vertical magnetic fields can in fact expel mass and angular momentum in an outflow from the upper and lower boundaries \citep{suz09,les13,bai13}.  However, the outflow rates have a strong dependence on simulation box size \citep{fro13}, and global simulations are really essential to address this form of transport.  Such global simulations, initialized with vertical magnetic fields, have been done \citep{suz14,ava16,zhu18,mis20,jac21}, but the wind is generally subdominant in terms of outward angular momentum transport compared to magnetic stresses within the disc, at least in the simulations of \citet{zhu18,mis20,jac21}. It is worth noting that vertical magnetic fields are necessary but not sufficient for the formation of magnetocentrifugal winds. These winds are commonly observed in isothermal local shearing box simulations with vertical magnetic fields but not in radiation magneto-hydrodynamic local shearing box simulations with vertical magnetic fields where thermal properties of the gas are calculated self-consistently \citep{Secunda+2024}. 

An interesting feature of simulations with vertical magnetic flux and winds is that they exhibit a tendency to spontaneously form ring-like concentrations of surface density very much like the expected outcome of the 
\citet{lig74} viscous instability.  Such ``zonal flows'', where radial pressure gradients balance departures from Keplerian rotation (i.e. geostrophic balance between the Coriolis and pressure gradient forces in the local co-rotating frame) have also been seen in shearing box simulations without net vertical magnetic flux and without winds \citep{joh09,sim12}.  However, the mechanism in that case appears to be associated with nonlinear self-organization in MRI turbulence \citep{rio19}.  In the presence of vertical magnetic fields, the fields end up concentrating in the low density ``gaps'' between the high mass density rings \citep{bai14,rio19,jac21}.  Possible explanations for this behavior are reconnection within vertical MRI channel modes that cause separation of mass from vertical flux \citep{bai14} or a true linear viscous instability caused by the winds \citep{rio19}.  Neither of these proposals appear to explain the simulation results of \citet{jac21}, who also note that the rings disappear above some critical threshold of vertical magnetization.     

\section{Modeling Outbursts of Cataclysmic Variable Stars, AM CVn Stars, Symbiotic Binaries, and Low-Mass X-ray Binaries}
\label{sec4}



The most compelling application of thermal/viscous instabilities in accretion discs to observational reality is in outbursting compact binaries.  There exist quite complete reviews describing the disc instability model of these outbursts, one recent \citep{Hameury0420} and two more ancient 
(\citealt{Osaki01996,Lasota0601}), so in the present section we will present and discuss the basics of the model, and supplemented only by the newest results and open questions.

\subsection{Dwarf novae}
\label{sec:dn}

Cataclysmic variable stars (CVs; see \citealt{Warner2003}) can be divided into two types: one consists of quasi-steady systems, known as ``nova-likes'' (NLs),
and the other consists of systems that are strongly variable on various timescales and are known as dwarf novae (DNs). It is now well established that the two
sets of binaries differ by the nature of their accretion discs. The membership of one of the two classes of CVs is determined by two
physical parameters: the mass-transfer rate $\dot M_{\rm tr}$ from the low-mass stellar companion of the white dwarf (WD) and the size $R_{\rm D}$ 
of the accretion disc around the WD.
These parameters correspond to observed quantities: luminosity and orbital period. The physical reason for the separation of CVs into
two classes is a thermal instability affecting accretion discs at temperatures below $\sim 10^4$ K. Such discs also suffer from a viscous
instability but since in geometrically thin discs the thermal time is much shorter than the viscous time, the disc variability is triggered
by thermal processes.

Linearized perturbations of the mass transfer and energy balance equations for a geometrically thin accretion disc by \citet{pir78} provided two necessary conditions for a stable disc. Consider an $\alpha$-disc, with a viscous heating rate $Q^+$ and a radiative cooling rate $Q^- = \sigma \, T_{\rm eff}^4 $. The thermal and viscous stability conditions  are, respectively:
\newcommand{\atc}{{a_{\rm T}}}
\renewcommand{\atc}{{k}}
\newcommand{\asigma}{{a_{\rm \Sigma}}}
\renewcommand{\asigma}{{l}}
\begin{equation}
\atc > 1  \, ;
\qquad
\cfrac{\atc-\asigma}{\atc-1} > 0  \,  
\qquad \mathrm{for}\quad
Q^- \propto T_{\rm c}^\atc \, \Sigma^\asigma\, ,
%
%
\label{eq.nec_cond_alpha_disc}
\end{equation}
where the dimensionless indices carry the dependence of the vertically integrated cooling rate $Q^-$ on the midplane temperature $T_{\rm c}$ and surface density $\Sigma$.
They are determined (sometimes only implicitly) after solving for the vertical structure and depend on the opacity and conditions at the surface\footnote{The second condition in equation \eqref{eq.nec_cond_alpha_disc} is equivalent to $a+1>0$ in equation \eqref{eq.Dnice}.}.

The thermal-viscous instability of $\alpha$-discs, when both conditions \eqref{eq.nec_cond_alpha_disc} are violated, is associated with hydrogen partial ionization. It is a result of (i) the  opacity in the outer layers of the disc at temperatures $\lesssim 10^4$~K dramatically drops and strongly depends on the temperature, $d \ln \kappa_{\rm R}/d \ln T \approx 7 - 10$; 
and (ii) 
the surface boundary condition dominates the structure~\citep{Faulkner_etal1983_1}. In particular, this is why irradiation can suppress instability by maintaining the temperature only at the disc surface.

\begin{figure}
    \centering
    \includegraphics[width=0.98\linewidth]{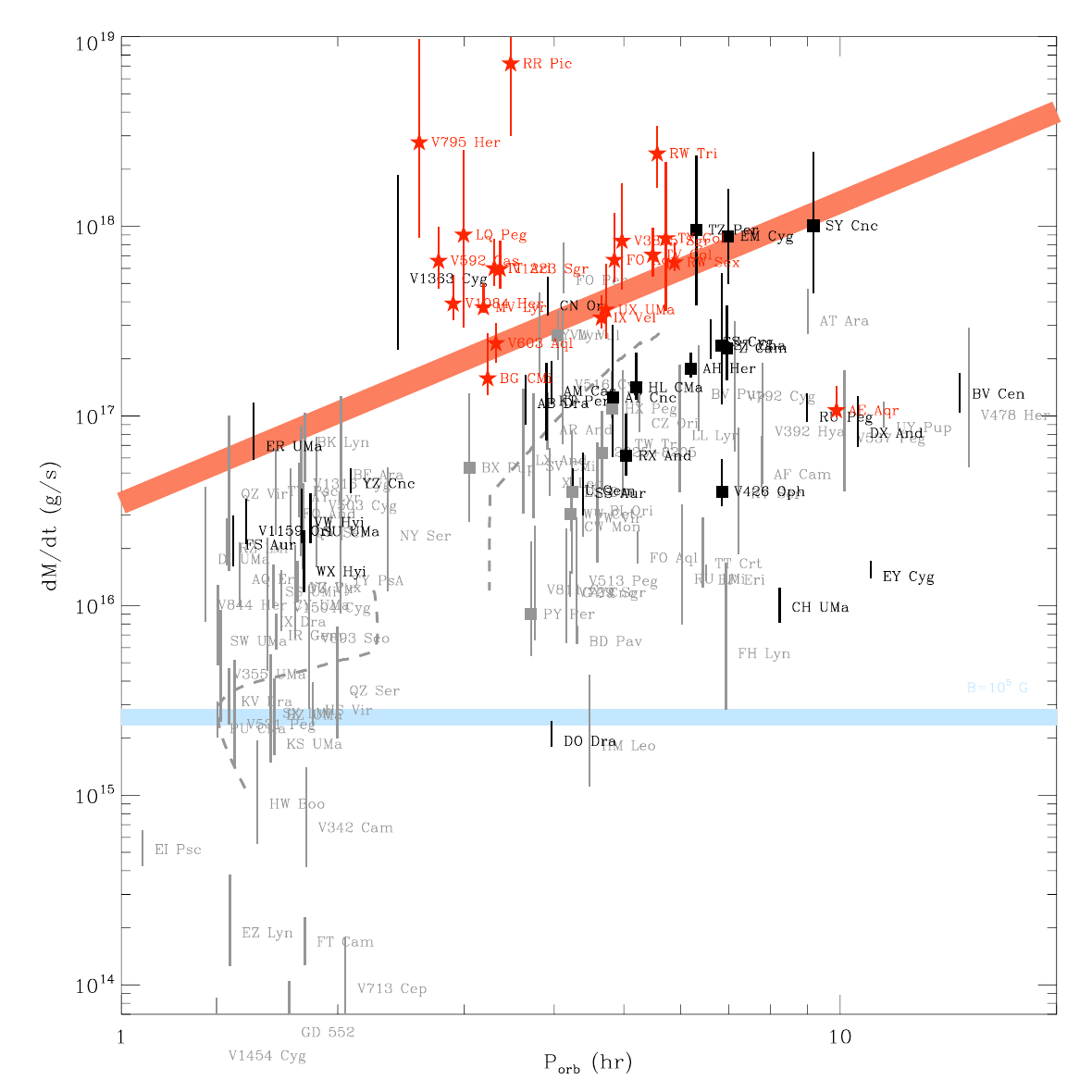}
    \caption{Mass transfer rates of CVs compared to the stability criterion Eq. (\ref{eq:stabCV}). Systems above
the upper (red) solid line are hot and stable. Systems below the lower (blue) line indicate
cold, stable discs if the white dwarf magnetic field $B \geq 10^5$ G. The dashed line represents the
expected secular mass transfer rate (\citealt{Knigge0611}). Square symbols indicate Z Cam type
dwarf novæ; (red) stars indicate nova-likes. Dwarf novae shown in black have a more complete
observed light-curve than those in grey.}
    \label{fig:stabCV}
\end{figure}
Since for a stationary Keplerian disc
\begin{equation}
    T_{\rm eff} \sim \left(\frac{\dot M}{R^{3}}\right)^{1/4},
\label{eq:statdisc}
\end{equation}
for every value of the accretion rate there exists a size of a disc at which it becomes unstable.
One could get the $\dot M_{\rm crit}(R_{\rm D})$ critical relation from Eq. (\ref{eq:statdisc}) 
by putting there $T_{\rm eff}=6500$ K (see, e.g., \citealt{Smak0182}), the temperature
corresponding to hydrogen recombination, but more detailed calculations give the relation \citep{Lasota0808}
\begin{equation}
    \dot{M}_{\text {crit }}^{+}(R)=8.07 \times 10^{15} R_{10}^{2.64} M_1^{-0.89} \mathrm{~g} \mathrm{~s}^{-1},
\end{equation}
where $R=R_{10} 10^{10}$cm and $M=M_1\Msun$.
A CV disc will be stable if
\begin{equation}
    \dot M_{\rm trans} > \dot{M}_{\text {crit }}^{+}(R_{\rm D})\, .
\label{eq:stabCV}
\end{equation}

The methods used to obtain reliable $\dot M$ from observed CV magnitudes are described in detail in \citet{Dubus0918}.
The disc radius is expressed through the orbital period by assuming that it is a fraction $f(q)$ of the binary separation $a$: $R_{\rm D} = f(q)a = 3.5\times 10^{10} f(q) M_1^{1/3} P^{2/3}_{\rm h}
\rm cm$,
where $q$ is the mass ratio (mass-of-the-companion/white-dwarf-mass) and $P_{\rm h}$ the orbital period in hours. In general, $f$ is well
approximated by $f = 0.6/(1 + q)^{2/3}$. The critical accretion rate can then be written as
\begin{equation}
    \dot{M}_{\text {crit }}\approx 3\times 10^{16} P_{\rm h}^{1.6} \rm g\,s^{-1}\, .
\label{eq:obscvs}
\end{equation}

Fig.~\ref{fig:stabCV} \citep{Dubus0918} exhibiting the $\dot M\left( P_{\rm orb}\right)$ relation deduced from observations of about 130~CVs shows clearly that the relation
Eq. (\ref{eq:obscvs}) separates these binary systems into two classes: those above are NLs while those below are DNs.

This confirms the basis of the dwarf-nova disc instability model (DIM): the outbursts are triggered by a thermal instability
due to hydrogen ionisation/recombination. However, the DIM must also reproduce faithfully the observed variety of DN outbursts. 
\begin{figure}
    \centering
    \includegraphics[width=0.68\linewidth]{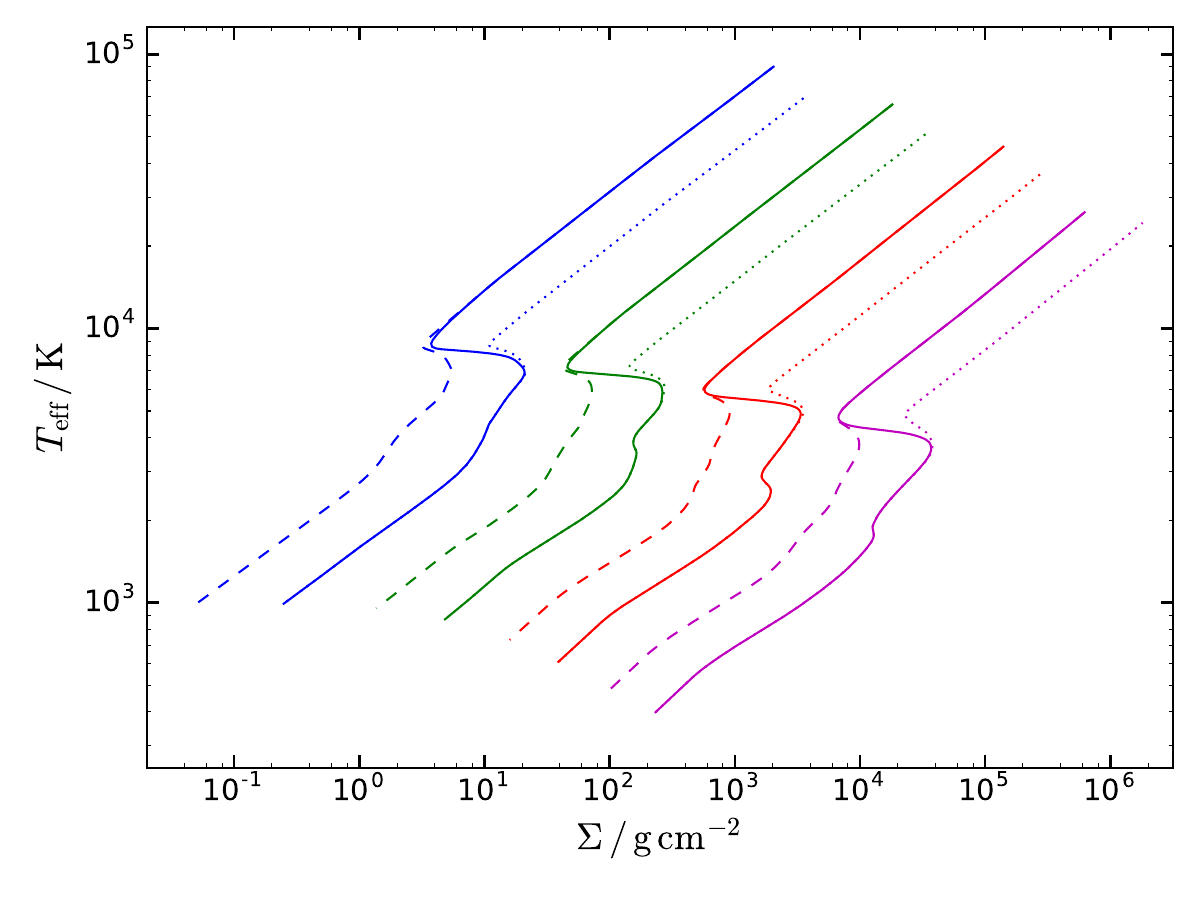}
    \caption{$\Sigma - T_{\rm eff}$ S-curves computed for a $1.35\Msun$ white dwarf are plotted
for various radii. The four sets of curves (each set includes a dashed, a solid
and a dotted line), from left to right, represent the S-curves at radii $R = 10^9, 10^{10},
10^{11}$  and $10^{12} \rm cm$, respectively. Solid lines represent the S-curves obtained
with temperature-dependent $\alpha$. Dashed and dotted lines represent S-curves
for $\alpha = 0.1$ and $\alpha = 0.01$ respectively \citep{Bollimpalli1218}.}
    \label{fig:scurves}
\end{figure}

\subsection{The dwarf-nova disc instability model}
\label{sec:dndim}
The thermal-viscous equilibrium states of accretion discs $Q^-= Q^+$, can be written as
\begin{equation}
    \sigma T_{\text {eff }}^4=\frac{9}{8} \nu \Sigma \omegak^2,
\end{equation}
where $\nu$ is the kinematic viscosity coefficient. Since $\nu = \alpha\, c_s^2\, \omegak^{-1} \propto T_{\rm c}$, where $c_s$ is
the isothermal speed of sound, solving the vertical energy transfer equation provides a relation between the effective and 
midplane temperatures. 
This allows one to represent the disc equilibria as a $T_{\rm eff} (\Sigma)$ or $Q (\Sigma)$ relation. 
As discovered
in the early 1980's this relation forms an S -- a necessary condition for the presence of outbursts forming a limit cycle \citep{Bath0482}.
The upper, hot branch of the S corresponds to disc states in outburst while during quiescence they are located on the lower, cold part
of the curve. 
The middle part of the S is thermally and viscously unstable. This is directly related to the fact that the slope of the curve $Q (\Sigma)$ equals the positive or negative number $(k-l)/(k-1)$, for the form of $Q^-$ suggested by \eqref{eq.nec_cond_alpha_disc}.

The states on both stable branches evolve on a viscous time and cross, up and down, the space between the two in a thermal time.
Since, in general, the DN outburst duration is shorter than the quiescence time, if the limit cycle is to reproduce the DN eruption cycle,
the viscous time on the upper branch of the S-curve should be shorter than the viscous time on the cold segment of the S.
Since the viscous time $t_{\rm vis} \sim T_c^{-1}$ this indeed is the case but as it has been quickly discovered, the difference
due to the temperature alone is not sufficient to produce a light curve resembling a usual DN outburst cycle \citep{Smak0184}.

The viscosity parameter $\alpha$ for hot discs has been determined from observations of DN outburst decays to be $\alpha_h = 0.1 - 0.2$ \citep{Smak0999,Kotko0912}. To reproduce the observed outbursts amplitudes, $\alpha_c$ in cold discs must be 5 to 10 times smaller.
These two conclusions are rather embarrassing from the theoretical point view because, until recently, MRI simulations with no net vertical flux usually produced $\alpha_h$ ten times smaller
and the value of $\alpha_c$ is difficult to estimate. These problems will be discussed in more detail in \S\ref{sec:mhd_dim}. Here we should stress that despite its shaky theoretical background the ``two - $\alpha$'s paradigm'' is very successful in reproducing many properties of the DN outbursts.

In practice this paradigm consists of joining, at a given radius, two S-curves into one ``effective'' S-curve with different values of $\alpha$
on the hot and cold branches~\citep[see, e.g.,][Eq. 17]{Hameury0485}. The results of such an operation are shown in Fig.~\ref{fig:scurves}.
\begin{figure}
    \centering
    \includegraphics[width=0.49\linewidth,height=0.48\linewidth]{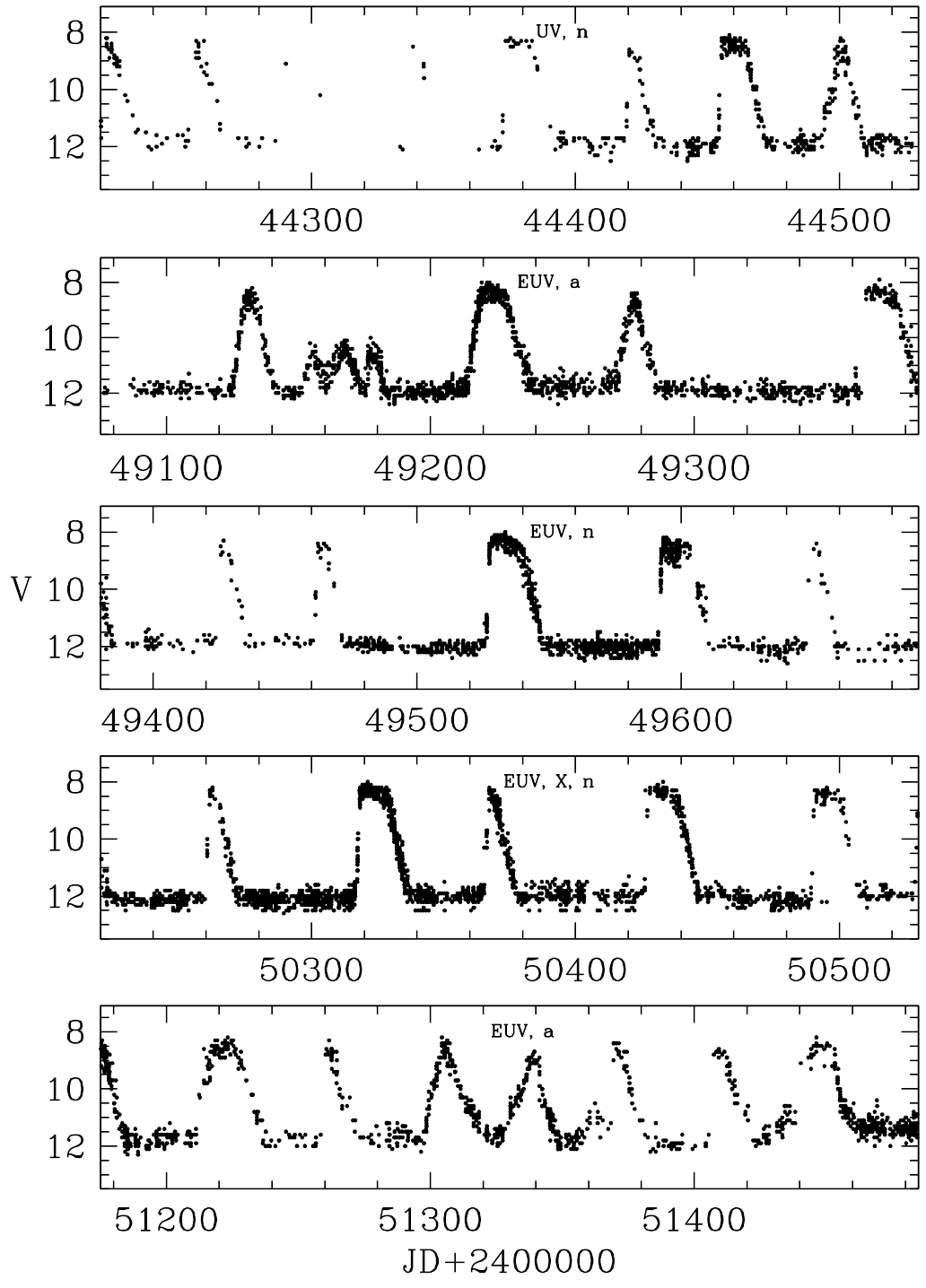}
    \includegraphics[width=0.49\linewidth]{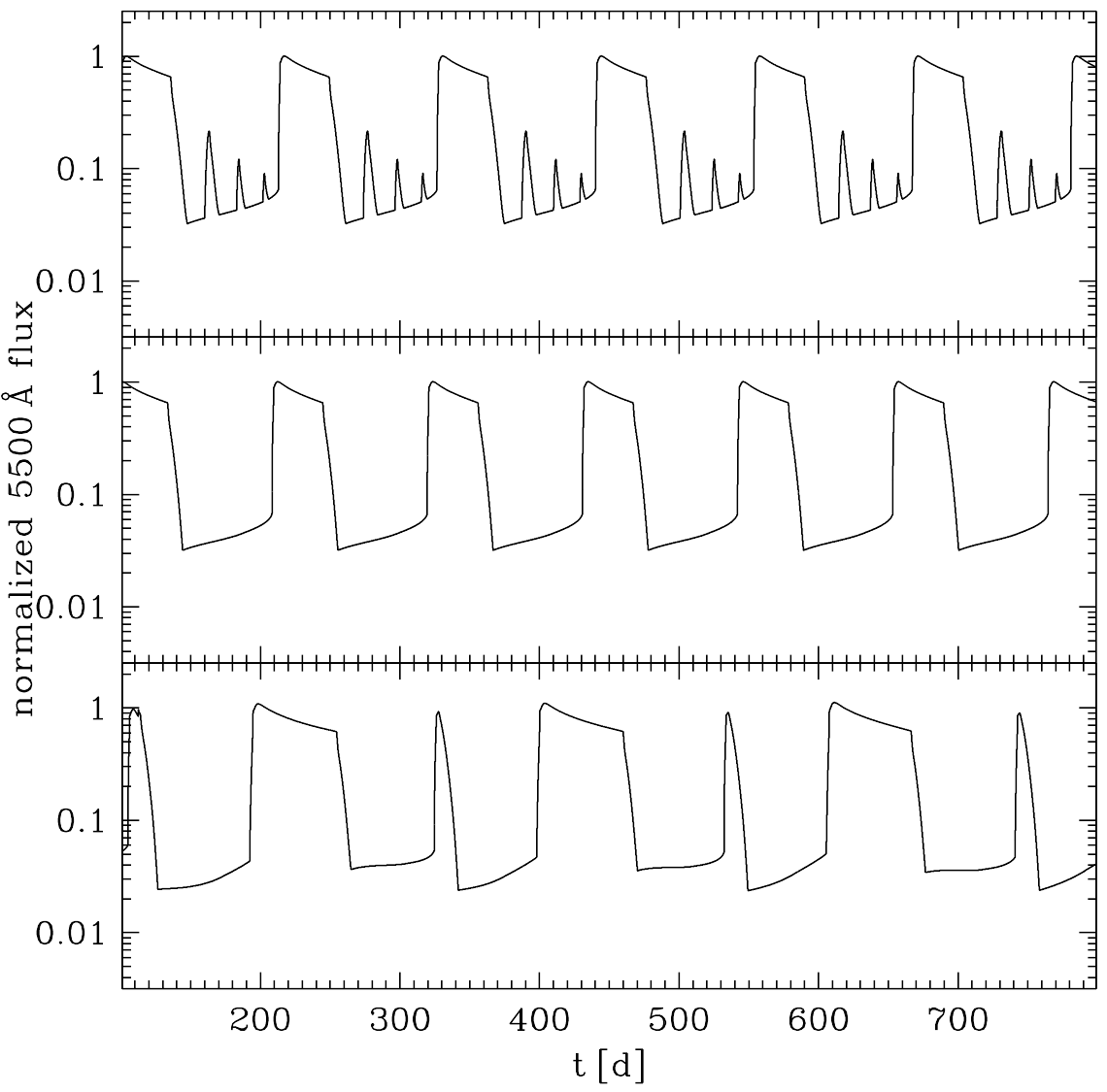}
    \caption{Left: the lightcurve of SS Cyg observed by the AAVSO; Right: three cases of model lightcurves \citep{Schreiber1003}
    of a binary with SS Cyg parameters. From the top: for $R_{\rm in}=R_{\rm WD}$ and $\dot M_{\rm tr}= const.$; 
    $R_{\rm in}>R_{\rm WD}$ and $\dot M_{\rm tr}= const.$, and $R_{\rm in}>R_{\rm WD}$ and $\dot M_{\rm tr}\neq const.$}
    \label{fig:SSCyg}
\end{figure}

Modifying $\alpha$ is, however, not sufficient to obtain model lightcurves resembling observations (see Fig.~\ref{fig:SSCyg}). One of the problems with the ``standard'' DIM is that
if the mass feeding rate is kept constant, outbursts form an exactly repetitive pattern never observed in reality. But there is no astrophysical reason for the mass transfer rate from the low mass secondary to be constant. On the contrary, this rate is observed to vary in practically all CVs. The most spectacular are VY Scl stars in which the mass-transfer rate is observed to stop for several months. Luminosity variations observed in polars must be caused by mass-transfer variations since there are no discs in these strongly magnetized binary systems. In eclipsing DNs one often observes strong brightness variations of the hot spot where the mass-transfer stream hits the outer disc. The standstills of Z Cam stars are naturally explained by modulation of the mass-transfer rate near the critical value for the disc instability. Of course this solution of the outburst periodicity problem is not fully satisfactory because the mechanism, or rather mechanisms, driving observed mass-transfer variation is unknown. Motions of magnetic spots near the L1 point have been invoked but no model makes specific predictions about the amplitudes or timescales of mass-transfer variation. Of course this is not supposed to be the part of the DIM but adds to it a free function.

\subsection{MHD Models of the Ionization Instability in Compact Binaries}
\label{sec:mhd_dim}

As we just discussed in the context of alpha-viscosities in the disc instability model, the outbursting behavior in cataclysmic variables provides the best observational constraints on the physical mechanisms of angular momentum transport in accretion discs.  Two $\alpha$'s are needed:  a high value $\sim0.1-0.2$ in the hot outburst state, and a much lower value in the cold quiescent state.  For hydrogen accreting systems, both hydrogen and helium are largely neutral in the quiescent state, and the dominant sources of free electrons are from alkali metals (particularly sodium) which have first ionization potentials less than that of hydrogen.  The abundance of these elements is rather low, however, implying that the plasma in the quiescent state would be too resistive to support robust MRI turbulence.  This alone might explain why $\alpha_c$ is so low in the quiescent state of hydrogen accreting systems, and might point to an alternative mechanism whereby it is the onset of good conductivity and MRI turbulence, not thermal instability per se, that drives outbursts \citep{gam98}.  Even if that were the case, however, there remains the problem of the high values of $\alpha_h$ in the outburst state, which are an order of magnitude larger than the values inferred from local shearing box simulations of MRI turbulence with no net vertical flux (\S\ref{sec:MRI}).  It may be that the discrepancy arises because of the limitations of the shearing box, and that more global magnetic field structures that can link different radii together are necessary to enhance the angular momentum transport \citep{tou96,kin07}.

\begin{figure}
    \centering
    \includegraphics[width=0.98\linewidth]{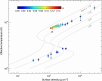}
    \caption{Stable thermal equilibria in vertically stratified shearing box simulations of MRI turbulence with diffusive radiation transport from \citep[from][]{hir14} and realistic opacities for a solar-composition plasma.  The colors indicate the time-averaged values of the Shakura-Sunyaev $\alpha$-parameter as measured from the MRI turbulent stresses and thermal pressures.  Grey curves show the thermal equilibrium predictions of the standard disc instability model with constant $\alpha$.}
    \label{fig:Hirose14_S-curve}
\end{figure}

Despite these issues, progress has been made by incorporating the thermodynamics associated with changes in opacity and cooling rates in shearing box simulations of MRI turbulence.  The first attempt at this \citep{lat12} used vertically unstratified shearing boxes with an optically thin cooling function which mocked up the expected cooling rates across the ionization transition.  Thermally stable equilibria at high and low temperatures were found, with unstable behavior in between, replicating the bistable behavior found in the disc instability model.  Vertically stratified shearing boxes with diffusive radiation transport and realistic grey opacities have also been done \citep{hir14,sce18a}, and found thermally stable equilibria at low and high temperatures \citep[as well as a stable intermediate temperature branch in the case of][]{sce18a}, with unstable behavior in between.  An example is shown in Fig.~\ref{fig:Hirose14_S-curve}.  In other words, simulations with MRI turbulence and realistic cooling are able to replicate stable branches of the S-curves used in the disc instability model.  Moreover, the vertically stratified simulations also find high values of alpha in the hot stable branch,
comparable to those inferred from observations, even without net vertical flux in these simulations.  The major difference with the standard disc instability model is that the enhanced values of alpha are only found near the low temperature end of the upper branch:  the alpha's return to lower values characteristic of standard zero net vertical flux simulations higher up on the upper branch.  The reason for the enhanced alpha's at the end of the branch are because the high opacities drive thermal convection, and this enhances the time-averaged MRI stresses (see \S\ref{sec:MRI}).  Some of the simulations by \citet{sce18a} incorporated resistivity and confirmed the conclusion of \citet{gam98} that MRI turbulence would be suppressed over much, but not all, of the lower branch --- the high surface density end of the branch can still be sufficiently conducting to sustain MRI turbulence.

Such MRI-based simulations of S-curves have also been done in the context of protoplanetary discs \citep[][with possible implications for FU~Ori outbursts]{hir15} and helium-dominated AM~CVn discs \citep{col18}, which exhibit both normal and superoutbursts.  Once again, convection enhances the effective $\alpha$ parameter at the lower end of the upper stable branch.  One interesting implication of the helium-dominated simulations is that convection can be persistent and still lead to enhanced $\alpha$ parameter near the end of the upper branch.  Moreover, because the first ionization potential of helium is so high, there are much more abundant sources of free electrons (particularly nitrogen) in the cool branch which lead to high enough electrical conductivity that ideal MHD should be a good approximation \citep{col18}.  It would therefore be interesting to explore the differences between AM~CVn outbursts and hydrogen-dominated dwarf novae as they might relate to these differences in conductivity in the quiescent state.

Whether merely enhancing the value of $\alpha$ at the low end of the upper branch is sufficient to explain outburst light curves is not as yet clear.  Ideally, one would like to run global radiation MHD simulations of MRI turbulence with realistic opacities to follow the evolution of heating and cooling fronts through an outburst cycle, but this remains numerically challenging.  An attempt was made to incorporate the variation of $\alpha$ on the upper branch into a dwarf nova outburst code \citep{col16} to compute outburst light curves, and these did replicate observed outburst and quiescent state time scales, as well as outburst amplitudes.  However, they also produced reflares in the decay to quiescence that are not generally observed.  A contributing factor to this was the poor modeling of the quiescent state \citep{col16}, where as we just discussed may in any case be too resistive to sustain MRI turbulence.  Much more work needs to be done to explore the physics of the quiescent state.

Radiation MHD simulations in vertically stratified shearing boxes with net vertical magnetic flux have also been pursued \citep{sce18b}.  Convection can again enhance $\alpha$ in the upper branch, but when the vertical magnetic field is strong enough, it can itself cause an increase in $\alpha$, and this enhancement can also occur in the quiescent state.  Vertical magnetic flux can also extract angular momentum in a magnetocentrifugal wind.  Incorporating such wind-driven transport into the disc instability model can replicate observed outburst lightcurves \citep{sce19,sce20}.

Before leaving this topic, it is also worth mentioning that, while global radiation MHD simulations of outbursts are not yet practicable, such simulations with MRI turbulence and no radiation transport have been pursued in Roche potentials.  These are shedding light on the relative roles of spiral waves and MRI turbulence in angular momentum transport \citep{ju16,ju17,pja20}, and how discs may spread outward in such potentials \citep{oya21,oha24}.  They have also recently been able to replicate the growth of eccentricity in the presence of MRI turbulence due to the 3:1 mean motion resonance \citep{oha24}, which is thought to be the origin of observed (positive) superhumps \citep{whi88,lub91}. 



\section{MHD Simulations and the Radiation Pressure Dominated Thermal/Viscous Instability}\label{sec5}


As we discussed in the introduction, thermal and viscous instabilities were first theoretically predicted to exist in the inner, radiation pressure and electron scattering-dominated region of the Shakura \& Sunyaev model (\citeyear{sha73}) of black hole accretion discs \citep{lig74,sha76}.  Standard accretion disc theory with radial advective cooling predicts an S-curve shape in the local accretion rate - surface density plane, with an upper stable slim disc branch and a lower stable gas pressure dominated branch \citep{che95}.  One would therefore expect limit cycle behavior just as in the hydrogen or helium ionization instabilities in cataclysmic variables and X-ray binaries, and indeed such outbursting behavior has been predicted \citep{hon91,szu98}.  However, observational evidence for the existence of radiation pressure dominated instability remains highly ambiguous.  Complex classes of flaring behavior are definitely present in black holes and neutron stars, e.g. GRS~1915+105 \citep{bel00}, IGR~J17091-3624 \citep{alt11}, the rapid burster MXB~1730-335 \citep{bag15}, V404~Cyg \citep{mot17}, the bursting pulsar GRO~J1744-28 \citep{cou18}, the ULX 4XMM~J111816.0-324910 in NGC 3621 \citep{mot20}, and Swift~J1858.6-0814 \citep{vin23}.
Where masses and distances are known, these all appear to be accreting at or above the Eddington limit, and it has been proposed that at least some of this flaring behavior is associated with some version of the radiation pressure dominated instability \citep{taa84,can96,bel97,taa97,nay00,jan00}.  However, there are also many black hole X-ray binaries that achieve high Eddington ratios in the soft state, whose spectra are well-fit by standard geometrically thin discs, and that show negligible variability in their soft X-ray emission \citep{gie04}.  Why, then, are they so stable?

A possible answer arose from early, vertically stratified shearing box simulations of MRI turbulence that incorporated diffusive radiation transport in the radiation pressure and electron scattering dominated regime \citep{hir09a}.  Surprisingly, these simulations were able to achieve steady thermal equilibria, even though the time-averaged stress appeared to approximately scale with total thermal pressure rather than having a dependency on the much smaller gas pressure \citep{hir09b}.  A cross-correlation analysis showed that the fluctuations in the thermal energy density responded to fluctuations in the turbulent energy with a lag of 5-15 local orbital periods, approximately the thermal time for these vertically stratified simulations.  Dissipation of turbulence into thermal energy requires some time, in contrast to the instantaneous relation that is assumed in the standard alpha prescription (\ref{eq:alpha_prescription}), and this was proposed as an explanation for thermal stability \citep{hir09a}.  The time-averaged stress still had an inverse relationship to surface density, however, so viscous instability was still a possibility \citep{hir09b}.

Unfortunately, this partial solution to the observed stability of the soft state was spurious.  Later attempts to replicate the simulated thermal stability failed, starting with vertically stratified shearing boxes \citep{jia13b}.  Even though the same stress to pressure time delay was present, these simulations were unable to maintain thermal equilibrium and always eventually underwent thermal runaway.  The simulated thermal stability in \citet{hir09a} was due in part to the use of flux-limited diffusion as a radiation transport algorithm and to a relatively small horizontal box size.  It is now known that such small horizontal domain sizes can dramatically affect the properties of MRI turbulence, and that the stresses have substantial large scale horizontal correlations \citep{sim12}.  Moreover, a scaling of stress with thermal pressure, which is what drives the thermal instability in the radiation pressure dominated regime, is only observed in gas pressure dominated shearing box simulations if the box size is larger than the thermal pressure scale height \citep{ros16}.  Runaway cooling in the radiation pressure dominated regime has also been observed in global general relativistic radiation MHD simulations \citep{sad16,mis16,lis22}.

That turbulent fluctuations lead pressure fluctuations due to dissipation is not surprising on short time scales, but if the alpha prescription is to have any validity in a causal sense, then presumably pressure fluctuations on longer time scales alter the saturation level of turbulence \citep{lat12,ros16}, and may even lead to stress fluctuations that lag pressure fluctuations \citep{ros17}.  Indeed, such a lag of order five orbital periods has been measured in vertically unstratified simulations with an optically thin cooling function that is artificially made to vary in time \citep{hel22a}. Strangely, this is smaller than the lag of pressure with respect to stress measured in optically thick vertically stratified simulations with radiation transport \citep{hir09a}, but this may be because the optically thin cooling in \citet{hel22a} responds instantaneously to changes in pressure, unlike diffusive cooling.  Measuring lags and leads as a function of fluctuation frequency, which can be done using Fourier methods \citep{now99}, might be fruitful in radiation MHD simulations.  The effects of simplified, fixed time delays on thermal stability have been investigated analytically by a number of authors.  A lag of stress relative to pressure can reduce the thermal instability growth rate, but not stabilize it \citep{lin11}, whereas a lag of pressure relative to stress can be stabilizing \citep{cie12}.  The stochasticity of turbulent fluctuations themselves could in principle stabilize the disc if the fluctuations are large enough \citep{jan12}, but simulations of MRI turbulence in a nominally thermally unstable situation find that, while thermal instability can be slowed, it is still generally present \citep{ros17}.

As we mentioned in \S\ref{sec:intro}, one possible explanation for thermal stability is that the inner parts of black hole accretion discs are supported against the vertical tidal gravity by magnetic, not thermal, pressure \citep{beg07,oda09}.  Indeed, simulations of cooling from a hot optically thin flow (the hard state) to a cooler flow (the soft state) found a stable, long-lived magnetically supported flow \citep{mac06}.  Global radiation MHD simulations of discs supported by magnetic pressure are found not to undergo thermal runaways \citep{sad16}, and discs that are initialized with magnetic field topologies that are able to achieve a magnetically dominated state also avoid thermal runaways \citep{mor18,mis22,lis22}.  Stabilization can also be achieved even if the midplane regions are supported by thermal pressure, but most of the accretion occurs at altitude in a magnetically dominated region:  radiation MHD simulations of the magnetically elevated disc flows discussed above in \S\ref{sec:meanfieldstresses} find that a radiation pressure dominated midplane can be stabilized in this regime \citep{lan19}. Such discs fit neither of the two categories of thin or slim, and the authors refer to them as ``puffy" discs.  Spectra of such discs have been computed in the mildly sub-Eddington regime, and resemble thermal spectra with a warm corona \citep{wie22}.

But this begs the question as to whether magnetically supported discs can have spectra that truly resemble the non-variable, strongly thermal spectra with no energetically significant corona that can exist in the high/soft state of black hole X-ray binaries \citep{dav06}.  Can a magnetically supported disc produce such thermal spectra, and can they show insignificant variability \citep{gie04}?  Note that observations by the Imaging X-ray Polarimetry Explorer might be able to constrain the presence of large scale coherent magnetic fields due to Faraday rotation \citep{bar24}, possibly providing a way of testing this solution to the observed lack of thermal instability in the high/soft state.

Yet another possible solution to the lack of observational evidence of thermal instability in the soft state is that modern simulations have not yet accurately captured the saturation behavior of the MRI under radiation pressure dominated conditions.  Indeed, it has only recently been convincingly demonstrated that MRI turbulence under {\it incompressible} conditions has an inertial range with constant energy flux from large scales to small scales \citep{kaw24}.  The simulation employed was a vertically unstratified shearing box, and even this required considerable computational resources.  The vertically stratified and global radiation MHD simulations that find thermal instability have nowhere near the dynamic range to explore this issue.  On the other hand, the photon diffusion scale is likely not much smaller than the driving scale in a radiation pressure dominated accretion disc \citep{tur03}, so that compressive damping is likely important on large scales, but incompressible fluctuations may still drive a cascade to smaller scales.   It may perhaps be that the saturation properties of MRI turbulence in radiation pressure dominated environments still requires a high dynamic range, and it might be worthwhile expending computational resources to investigate this problem.

While our discussion here has focused on X-ray binaries, accretion discs in luminous active galactic nuclei (AGN) are expected to be even more radiation pressure dominated, so one might ask how thermal/viscous instabilities might manifest there.  A major complication is that the fiducial temperatures are much lower (ultraviolet) in AGN compared to X-ray binaries, and simply on the basis of the free-free opacity therefore being larger, AGN discs were predicted early on to be thermally stable \citep{pri76}. However, vertically stratified shearing box simulations with just free-free and electron scattering opacities are apparently unable to achieve long-lived thermal equilibria \citep{jia16}.  On the other hand, additional opacity sources are also present.  AGN discs have densities and temperatures similar to those in the envelopes of massive stars \citep{jia16}, and iron can significantly enhance the Rosseland mean opacity $\kappa$ over that of pure electron scattering $\kappa_\mathrm{T}$ near temperatures $\sim2\times10^5$~K \citep{jia15}.  Vertically stratified shearing box simulations {\it are} able to achieve long-lived thermal equilibria when this opacity enhancement is present \citep{jia16}.  In the standard model, where radiation pressure provides hydrostatic support against the vertical tidal gravity, the iron opacity region is located at radius given by
\begin{equation}
    \frac{r}{r_{\rm g}}=\left(\frac{4c^8}{G^2M^2\alpha^2 \kappa^2a^2T^8}\right)^{1/3}=9.6\alpha^{-2/3}\left(\frac{\kappa}{\kappa_{\rm T}}\right)^{-2/3}\left(\frac{T}{2\times10^5{\rm K}}\right)^{-8/3}\left(\frac{M}{10^8M_\odot}\right)^{-2/3},
\end{equation}
where $r_g=GM/c^2$ is the gravitational radius of the black hole.
With the chosen black hole mass, this ranges from $44 r_{\rm g}$ ($\alpha=0.1$) to $210 r_{\rm g}$ ($\alpha=0.01$), depending on the Shakura-Sunyaev $\alpha$ parameter.   The corresponding effective temperature at the photosphere at this radius is
\begin{equation}
    T_{\rm e}=9800\,\,{\rm K}\left(\frac{\dot{M}}{\eta\dot{M}_{\rm Edd}}\right)^{1/4}\alpha^{1/2}\left(\frac{\kappa}{\kappa_{\rm T}}\right)^{1/2}\left(\frac{T}{2\times10^5{\rm K}}\right)^2\left(\frac{M}{10^8M_\odot}\right)^{1/4}.
\end{equation}
The thermal time scale is
\begin{equation}
    t_\mathrm{th}\sim(\alpha\Omega)^{-1}=\frac{2c}{\alpha^2\kappa aT^4}=4.6\times10^{-4}\alpha^{-2}\left(\frac{\kappa}{\kappa_{\rm T}}\right)^{-1}
\end{equation}
and ranges from 17 days ($\alpha=0.1$) to 4.6 years ($\alpha=0.01$).

Global simulations find, indeed, that thermal runaways do not occur in this radial range in radiation pressure dominated flows \citep{jia20}, see Fig.~\ref{fig:AGNIron}.  However, considerable variability is still present.  The iron opacity peak causes vertical density inversions which are convectively unstable, and this leads to convective/radiative cyclic behavior with variable enhancements of MRI turbulent stresses.  These in turn cause radial concentrations of surface mass density that appear and disappear.  Strong luminosity variability is associated with these cycles, with a characteristic time scale of order a year, consistent with the thermal time.  This might therefore be a possible explanation for the characteristic time scale inferred in damped random walk models of AGN optical variability \citep{bur21}.

Away from the iron opacity dominated region, thermal stability can again be achieved if the disc is supported vertically by magnetic pressure \citep{jia19,cur23}.    In addition to coherent and turbulent Maxwell stresses, one such simulation had substantial radiation viscosity:  the off-diagonal terms in the radiation stress tensor in the optically thin layers drove more than half the accretion rate \citep{jia19}.

As noted in \S\ref{sec:meanfieldstresses}, a recent cosmological zoom-in simulation has produced a very strongly magnetized accretion disc on scales of order 100 gravitational radii \citep{hop24a,hop24b}.  Such a disc is much less dense and less gravitationally unstable than a standard Shakura-Sunyaev disc.  Analytic scalings based on the simulation results \citep{hop24c} suggest that radiation pressure will become more important near the black hole at near or super-Eddington accretion rates.  A very recent extension of these simulations to several gravitational radii around the black hole does indeed find that radiation pressure is comparable, but does not dominate, magnetic and turbulent pressure \citep{hop25}.  These simulations generally have much higher inflow speeds and lower surface mass densities and vertical optical depths than both standard Shakura-Sunyaev models \citep{hop24c} and magnetically elevated discs that are initialized with tori around the black hole.  Whether this is consistent with observational constraints remains to be seen, but it appears likely that there remains a large parameter space of strongly magnetized accretion discs that has yet to be explored.

Finally, before leaving this section we should point out that there are also now many global simulations of super-Eddington accretion onto black holes and neutron stars \citep{jia14,sad14,mck14,sadnar16,ino24}, which may be of relevance to the highest luminosity sources, e.g. ultra-luminous X-ray sources.  Thermal
stability is less of an issue in this regime, and instead the focus has been on the efficacy of photon trapping and radial advection of heat, compared to vertical diffusive and advective escape.  Outflows driven by continuum radiation pressure are also an important issue here. We refer the interested reader to the cited references for more information.

\section{Conclusion:  Outstanding Questions and Directions for Future Research}\label{sec6}

\begin{enumerate}
\item{Thermal and viscous instabilities are most evident in outbursting cataclysmic variables and X-ray binaries, and these systems still provide the strongest constraints on the angular momentum transport mechanism that operates in accretion discs in this regime.  As we discussed here, the $\alpha$-based disc instability model, with some ad hoc tweaks, can reproduce observed outburst light curves.  Some aspects of these light curves can also be reproduced with models based on local simulations of MRI turbulence, but more work is needed to investigate and constrain magnetic angular momentum transport in these systems.  In particular, simulations that allow for the possibility of magnetocentrifugal winds, and that also address the physics of the quiescent state, are needed.  The simulations that do exist demonstrate that composition matters:  hydrogen and helium discs are predicted to behave very differently, and more work is needed to see how this might manifest observationally.}

\item{The original predicted thermal/viscous instability in the radiation pressure and Thomson scattering dominated inner zone of the classical Shakura-Sunyaev alpha disc model does not appear to be present in the high/soft state of black hole X-ray binaries.  Spectra in this state are well-fit by simple multi-temperature blackbody disc models, and the variability in this state is extremely small.  Currently it seems likely that the solution to this problem of observed stability involves some form of magnetic pressure support in the disc, and how one achieves this while (a) still maintaining a thermal spectrum, (b) still maintaining minimal variability, and (c) not Faraday depolarizing the observed polarization signal, remains an outstanding research problem.}

\item{More global simulations that incorporate the coupling between magnetic angular momentum transport mechanisms and the thermodynamics (heating and cooling) of the disc are needed, in all astrophysical contexts from cataclysmic variables to active galactic nuclei.  There is a rich interplay that is only just beginning to be elucidated.  Particularly important is how thermodynamics couples not only to weak field MRI turbulence but also to transport by large scale magnetic field stresses, including magnetocentrifugal winds. How angular momentum transport is coupled to thermodynamics is also extremely important in understanding the propagation of heating and cooling fronts in the disk during the dwarf nova outburst cycle, though this is likely to be very challenging given current computational resources.}

\item{At the same time, local shearing box experiments that are carefully constructed to address basic physics questions still need to be pursued.  On what fluctuation time scales do stresses lead and lag thermal pressure fluctuations?  How does field topology and magnetic Prandtl number affect turbulent stresses, especially when coupled with thermodynamics?  Can the recent successful capture of an inertial range in {\it incompressible} MRI turbulence \citep{kaw24} be generalized to compressive MRI turbulence in a radiation pressure dominated plasma?  How does thermodynamics couple to the dynamo that maintains a strongly magnetized turbulence state?  While such local experiments neglect the radial coupling due to radial flows and large scale magnetic fields, answers to these fundamental questions would still be of great use.}

\item{Finally, in addition to theoretical work, more observational constraints on the presence and strength of turbulence in different astrophysical discs are desperately needed.  
In addition to, e.g., modeling light curves, useful information can be obtained by measuring the turbulent broadening of spectral lines in protoplanetary discs and AGN discs~\citep{Flaherty+2018,Flaherty+2024,Ochman+2024}, or even in binaries with  
compact stars~\citep{Horne1995,Skidmore+2000}.  In addition, imaging
observations of rings in protoplanetary discs~\citep{Rosotti+2020} provide constraints in that context.  Direct constraints on the strength of magnetic fields in discs can also be obtained through detections of polarized disc emission \citep{bar24}.}
\end{enumerate}

\backmatter



\bmhead{Acknowledgements}

We thank the referees for comments that helped improve this paper, and the International Space Science Institute in Bern for hosting a productive workshop on the first 50 years of research on accretion discs.  We also want to acknowledge the tremendous scientific career contributions of our dear friend and colleague Jean-Pierre Lasota, who unfortunately passed away while this review was being written.


\section*{Declarations}

Conflict of interest or competing  interests --- not applicable.









\bibliography{main,time-dep-sol, Lasota_DIM}

\end{document}